\documentclass[lettersize,journal]{IEEEtran}

\usepackage{amsthm}
\usepackage{cite}
\usepackage{amsmath}
\usepackage{amssymb}
\usepackage{amsfonts}
\usepackage{amsthm}
\usepackage{graphicx}
\usepackage{fancyhdr}
\usepackage{svg}
\usepackage{bbm}
\usepackage{textcomp}
\usepackage{xcolor}
\usepackage{algorithmicx,algpseudocode}
\usepackage[linesnumbered,ruled]{algorithm2e}
\usepackage{booktabs}
\usepackage{bbold}
\usepackage{kotex}
\usepackage{mwe}
\usepackage{multicol}
\usepackage{multirow}
\usepackage[normalem]{ulem}
\usepackage[bottom]{footmisc}
\usepackage{mwe}
\usepackage{color}
\usepackage{booktabs,multirow}
\usepackage{tabularx}
\usepackage{tikz}
\usepackage{kotex}
\usetikzlibrary{patterns,arrows}
\usepackage{physics}
\usepackage{verbatim}
\usepackage{subcaption}
\usepackage{pifont}
\usepackage[bottom]{footmisc}
\usepackage{circledsteps}
\usepackage{microtype}
\usepackage[export]{adjustbox}
\usepackage{enumitem}

\newcounter{descriptcount}
\newlist{enumdescript}{description}{1}
\setlist[enumdescript,1]{
before={\setcounter{descriptcount}{0}
\renewcommand*\thedescriptcount{\arabic{descriptcount}}},font={\bfseries\stepcounter{descriptcount}Q\thedescriptcount:}
}

\usepackage[nolist]{acronym}

\SetKwRepeat{Do}{do}{while}
\SetKwInOut{Input}{Input}
\SetKwInOut{Output}{Output}

\theoremstyle{definition}

\theoremstyle{Theorem}

\definecolor{color1}{RGB}{228,26,28}
\definecolor{color2}{RGB}{55,126,184}
\definecolor{color3}{RGB}{77,175,74}
\definecolor{color4}{RGB}{152,78,163}
\definecolor{color5}{RGB}{255,127,0}
\definecolor{color6}{RGB}{241,241,241}
\definecolor{color7}{RGB}{156,156,156}
\definecolor{color8}{RGB}{96,96,96}
\definecolor{color0}{RGB}{162, 20, 47}

\newcommand{\eg}{{\sl e.g., }}
\newcommand{\ie}{{\sl i.e., }}

\newcommand{\tblue}[1]{{\color{black}{#1}}}

\newcommand{\tpurp}[1]{{\color{black}{#1}}}

\newcommand{\cmark}{\ding{51}}
\newcommand{\xmark}{\ding{55}}

\newcommand{\BfPara}[1]{{\noindent\bf#1.}\xspace}

\title{\fontsize{16.5}{24}\selectfont Integrating Pre-Trained Language Model with Physical Layer Communications}

\author{
    Ju-Hyung Lee \ 
    Dong-Ho Lee  \ 
    Joohan Lee \
    Jay Pujara 
    \thanks{    
    J.-H. Lee and J. Lee are with Ming Hsieh Department of Electrical and Computer Engineering, University of Southern California, Los Angeles, USA (Emails: \{juhyung.lee, joohanl\}@usc.edu).}
    \thanks{D.-H. Lee and J. Pujara are with Information Science Institute, University of Southern California, Marina Del Rey, USA (Emails: \{dongho.lee, jpujara\}@usc.edu).}
}

\begin{document}
\maketitle
\begin{acronym}
 \acro{AWGN}{additive white Gaussian noise}
 \acro{BER}{bit error rate}
 \acro{SER}{symbol error rate}
 \acro{DL}{deep learning}
 \acro{GPU}{graphic processing unit}
 \acro{LoS}{line-of-sight}
 \acro{NLoS}{non-LOS}
 \acro{MIMO}{multiple-input multiple-output}
 \acro{MLP}{multilayer perceptron}
 \acro{MSE}{mean squared error}
 \acro{NN}{neural network}
 \acro{QAM}{quadrature amplitude modulation}
 \acro{SNR}{signal-to-noise ratio}	
 \acro{FEC}{forward error correction}
 \acro{LDPC}{low-density parity-check}
 \acro{TX}{transmitter}
 \acro{RX}{receiver}
 \acro{DNN}{deep neural network}
 \acro{CE}{cross entropy}
 \acro{i.i.d.}{independent and identically distributed}
 \acro{AI}{artificial intelligence}
 \acro{ML}{machine learning}
 \acro{LLM}{large language model}
 \acro{LM}{language model}
 \acro{OFDM}{orthogonal frequency-division multiplexing}
 \acro{OOD}{out-of-domain}
 \acro{OOV}{out-of-vocabulary}
 \acro{NLP}{natural language processing}
 \acro{PHY}{physical layer}
 \acro{E2E}{end-to-end}
 \acro{VQ-VAE}{vector quantised variational auto encoder}
 \acro{HARQ}{hybrid automatic repeat request}
 \acro{ARQ}{automatic repeat request} 
 \acro{JSCC}{joint source-and-channel coding} 
 \acro{CDL}{clustered delay line}
\end{acronym}

\begin{abstract}
The burgeoning field of on-device AI communication, where devices exchange information directly through embedded foundation models, such as language models (LMs), requires robust, efficient, and generalizable communication frameworks. However, integrating these frameworks with existing wireless systems and effectively managing noise and bit errors pose significant challenges.
In this work, we introduce a practical on-device AI communication framework, integrated with physical layer (PHY) communication functions, demonstrated through its performance on a link-level simulator.
Our framework incorporates end-to-end training with channel noise to enhance resilience, incorporates vector quantized variational autoencoders (VQ-VAE) for efficient and robust communication, and utilizes pre-trained encoder-decoder transformers for improved generalization capabilities.
Simulations, across various communication scenarios, reveal that our framework achieves a $50 \%$ reduction in transmission size while demonstrating substantial generalization ability and noise robustness under standardized 3GPP channel models.


\end{abstract}

\IEEEpeerreviewmaketitle

\section{Introduction} \label{sec:intro}

The increasing capabilities of mobile devices and the advancements in \ac{LLM}, particularly foundation models, have paved the way for a new era of on-device \ac{AI}.\footnote{Throughout this paper, we will use the term ``on-device AI" and ``on-device AI/\ac{LM}" interchangeably to refer to a foundation model or a language model deployed at both \ac{TX} side and \ac{RX} side in a distributed manner.} This paradigm shift allows devices to possess their own AI capabilities, enabling tasks such as real-time translation, personalized recommendations, and even autonomous driving, all while ensuring privacy and efficiency \cite{dhar2021survey, bommasani2022opportunities, shijie2022recommendation, cui2023survey}.

On-device AI communication enables devices equipped with advanced foundation models like \ac{LM}s to directly exchange information, forming a network of distributed intelligence. This distributed AI-to-AI communication requires efficient and accurate transmission of information between devices. \tblue{The ``encoded representation" generated by the AI/LLM at the \ac{TX} side (called \emph{AI-Src-Enc}) needs to be reliably transmitted and accurately interpreted as "decoding input" by the AI/LLM at the \ac{RX} side (called \emph{AI-Src-Dec}) \cite{deletang2023language}.}

However, seamlessly integrating this novel technology with existing communication infrastructure presents a significant challenge. Maintaining backward compatibility is a crucial requirement in wireless communication systems, allowing devices designed for older networks (\eg 4G and 5G) to operate seamlessly with newer technologies (\eg 6G). This interoperability ensures a smooth transition between networks, such as handover and roaming, enhancing user experience. 
For on-device AI communication, ensuring backward compatibility is still essential, enabling existing communication infrastructure to support this new paradigm without requiring major modifications. Thus, it is required to integrate the on-device AI model application into wireless communication systems, ensuring they adhere to existing legacy protocols for seamless interoperability \cite{lin2022overview}.

Furthermore, effective on-device AI communication demands several key qualities: (1) \textit{Robustness}: The system must be resilient to communication errors (\eg bit error) and noise (\eg channel fading) inherent in dynamic wireless environments; (2) \textit{Efficiency}: The system must efficiently compress and transmit data while maintaining accuracy and fidelity; and (3) \textit{Generalization Capability}: The system should be able to handle diverse inputs and adapt to various communication scenarios.
Meeting these requirements presents another challenge, requiring novel approaches in system design and optimization.

\subsection{Related Works}

\BfPara{\tpurp{Semantic communication}} \quad
Semantic communication, which focuses on effectively transmitting semantic (meaningful) information rather than merely receiving individual symbols or bits~\cite{qin2021semantic}, is the most relevant research direction for distributed AI-to-AI communications. 
In order to accomplish this, semantic communication systems leverage \ac{NN} at both the \ac{TX} and \ac{RX} to extract and decode the semantic information.
Semantic communication research can be classified based on the types of data they utilize, including image~\cite{bourtsoulatze2019deep, shao2021learning, huang2022toward}, video~\cite{tung2022deepwive, wang2022wireless, jiang2022wireless}, and speech/text data~\cite{xie2021deep, han2022semantic, weng2023deep}.
These works pioneered the inclusion of \ac{AI} in source coding within transceivers, optimizing \ac{JSCC} under specific noisy channels and data distributions.

However, such semantic communication framework poses significant challenges for practical application in on-device AI communication. Firstly, it demands major modifications to existing legacy communication protocols (\eg 5G-NR protocol), distinguishing it from practical distributed AI-to-AI communication systems. \ac{JSCC} necessitates that both channel coding and communication modules be modified and trainable, requiring major modifications in existing legacy communication protocols that are impractical in conventional wireless communication systems. 

Secondly, much of the existing research assumes ideal channel conditions (e.g., \ac{AWGN}) and bases analysis on information theory and symbol-level transmission. This overlooks the complexities of non-ideal (realistic) channel conditions and the necessity for link (or system)-level analysis and practical bit-level transmission.

Advancing the pioneering efforts in semantic communication towards practical application in on-device AI communication necessitates two critical components:
(1) \emph{Reliable} communication systems need to seamlessly operate with existing infrastructure. This typically involves interacting with various physical-layer communication function modules like mappers, \ac{FEC}, and \ac{OFDM}, each serving a distinct purpose. Therefore, it becomes necessary to integrate AI-Src-Enc and AI-Src-Dec within the transceiver system, with the operational parameters of modules compatible with, or analogous to, those in 5G-NR physical-layer modules.
(2) \emph{Efficient} communication systems should be able to compress the output of AI-Src-Enc and accurately decode the input of AI-Src-Dec. Developing an effective compression approach is crucial for reducing the size of semantic messages while retaining their essential information.

\begin{figure*}
\centering
\begin{subfigure}[t]{.7\linewidth}
    \centering  
    \includegraphics[width=\linewidth]{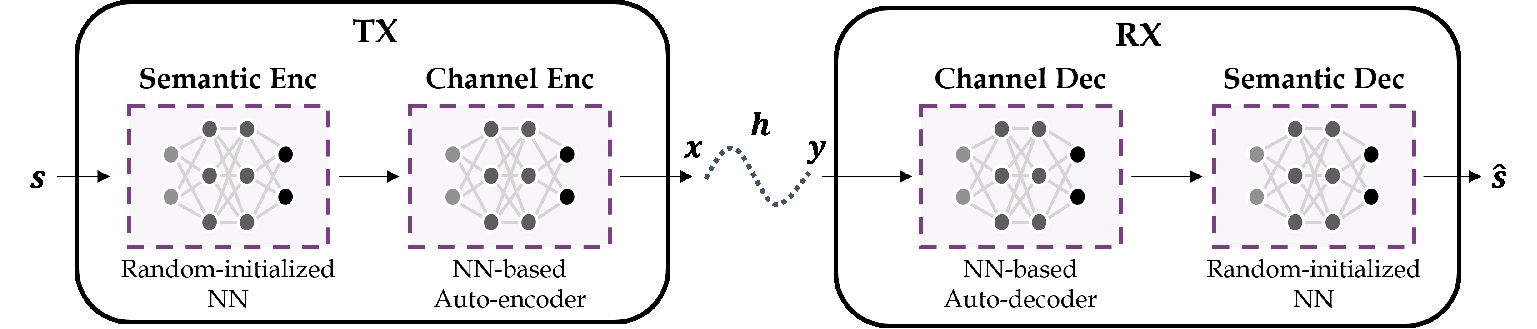}
    \caption{Semantic communication system.}
    \label{fig:overview_conventional}
\end{subfigure} \\ \vspace*{.015\textwidth}
\begin{subfigure}[t]{.7\linewidth}
    \centering
    \includegraphics[width=\linewidth]{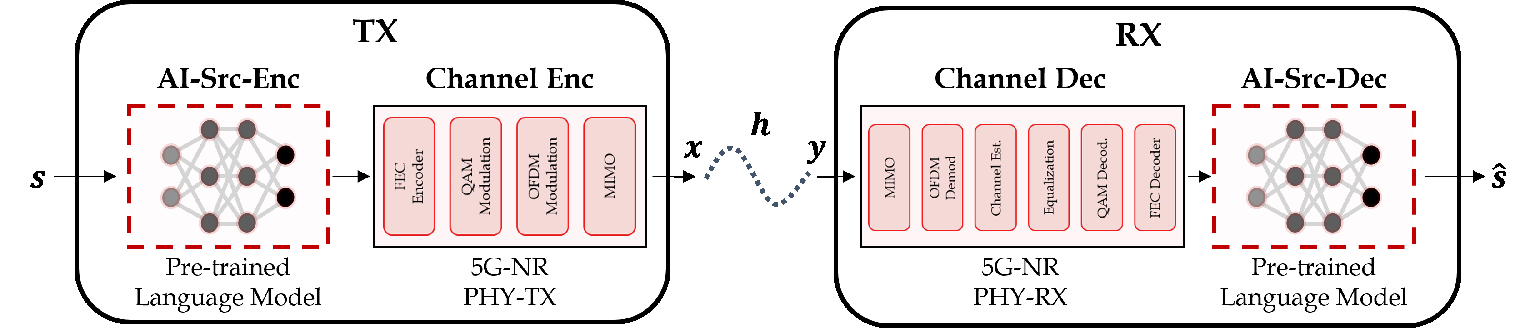}
    \caption{On-device AI/LM communication system.}
    \label{fig:overview_proposed}
\end{subfigure}
\caption{Overview of semantic communication systems and on-device AI/LM communication system.}
\label{fig:overview}
\end{figure*}

Fig.~\ref{fig:overview} shows the difference between semantic communication and on-device AI communication systems.

\BfPara{Pre-Trained Language Model} \quad
Recent studies on text semantic communication exploit transformer architectures~\cite{vaswani2017attention} to extract semantics at the \ac{TX} and recover the original information at the \ac{RX}~\cite{Semantic_Geoffrey2021, hu2022one}.
However, such frameworks may have the following challenges:
(i) training an \ac{E2E} semantic communication pipeline requires a huge computational effort due to the many randomly initialized parameters of semantic encoder/decoder to be trained on; (ii) difficulty in handling \ac{OOV} since they only use a set of whitespace-separated tokens in the training data.
\emph{General-purpose} communication systems should be able to effectively handle any data, including \ac{OOD} data that lies outside the training data used to develop the framework. This necessitates using pre-trained language models with superior generalizability, allowing them to perform well even when encountering unforeseen data variations.

Pre-trained language models, which have been extensively trained on vast web datasets, are revolutionizing the field of \ac{NLP}~\cite{brown2020language, srivastava2022beyond}.
These models, often referred to as ``foundation models'', possess the ability to effectively generalize even to data where they have not specifically been trained for~\cite{bommasani2022opportunities}.
Their capabilities are a result of leveraging transfer learning and scaling techniques.
Transfer learning involves utilizing knowledge acquired from one task and applying it to another~\cite{thrun1998lifelong}.
In the context of deep learning, pre-training is the prevailing approach to transfer learning, where a model is initially trained on a surrogate task and subsequently fine-tuned to adapt to the downstream task of interest.

\subsection{Contributions}
To design and evaluate a practical and widely applicable on-device AI communication framework, we focus on three key questions:
\begin{enumdescript}
    \item \vspace{0mm} \textit{How can we integrate a pre-trained language model in practical \ac{PHY} communication systems and evaluate its reliability in realistic network scenario?}
    \item \vspace{0mm} \textit{How can we efficiently compress the output of the AI-Src-Enc and accurately decode the noisy input of the RX AI-Src-Dec?}
    \item \vspace{0mm} \textit{How can we leverage the knowledge gained from a large dataset to build a general-purpose encoder/decoder for system-wide application?}
\end{enumdescript}

Our main contributions, which address these challenges, are summarized as follows:
\begin{itemize}
\item  \vspace{0mm} 
We integrate a language model with the link-level simulator, \texttt{NVIDIA Sionna}~\cite{hoydis2022sionna}, incorporating  5G-NR \ac{PHY} communication functions (\eg Polar channel coding and QAM mapper). We evaluate its performance on a channel that contains not only noise but also delay dispersion (\eg 3GPP CDL-family channel), seeing the efficacy of on-device AI communication. 
Furthermore, we propose a \emph{noise-tuning} method to optimize its reliability. This addresses \textbf{Q1}.

\item  \vspace{0mm}
Converting on-device AI information (\eg output of AI-Src-Enc and input of AI-Src-Dec) into bit-level information often increases data load.
Additionally, even minor bit errors, induced by channel noise, can greatly hinder communication.
Addressing this, we propose a novel approach that leverages vector quantization techniques~\cite{van2017neural} for efficient compression and decoding of on-device AI information.
This approach transforms high-dimensional vectors into discrete data via a codebook, significantly reducing transmission overhead even while mitigating the impact of bit errors (\ie robustness against channel noise). 
This addresses \textbf{Q2}.

\item  \vspace{0mm} 
We incorporate encoder-decoder transformers into \ac{E2E} on-device AI communication systems, initializing them with pre-trained weights. This approach reduces the reliance on channel-specific and data distribution-specific optimization, thus enhancing the system's generalizability; This addresses \textbf{Q3}.
In particular, we employ a pre-trained encoder-decoder transformer (\texttt{BART}~\cite{lewis-etal-2020-bart}) to initialize the parameters of AI-Src-Enc/Dec so that the pipeline itself requires little or no computational effort and use a pre-trained tokenizer to effectively handle \ac{OOV} so that our pipeline can be generalized to any other text.

\item We release source code for the experiments to promote reproducible ML research in wireless communication.\footnote{https://github.com/abman23/on-device-ai-comm}
\end{itemize}

\subsection{Paper Organization and Notation}
This paper is organized as follows:
Sec.\ref{sec:background} provides the background information on two key concepts: \emph{on-device AI communication} and \emph{\ac{VQ-VAE}}
After presenting a detailed system model for the on-device AI communication systems in Sec.\ref{sec:systemmodel}, 
Sec.~\ref{sec:exp} provides comparison results and ablation studies to assess the contribution of each of our proposed approaches, with a specific emphasis on the accuracy and efficiency of text transmission.
Then, we also present an in-depth analysis of our proposed \emph{noise-tuning} methods, followed by concluding remarks in Sec.~\ref{sec:conclusion}.

\BfPara{Notation} \quad
Random variables are denoted by capital italic font, \textit{e.g.,} $X ,Y$, with realizations $x, y$, respectively.
$I(X; Y)$, $p(y|x)$ and $p(x,y)$ represent the mutual information, conditional probability, and joint probability distribution of the two random variables $X$ and $Y$. Multivariate random variables are represented with capital bold font, \textit{e.g.,} $\mathbf{Y} = [Y_0, Y_1]^T$. Vectors are represented using a lowercase bold font, \textit{e.g.,} $\mathbf{y}$. We use $\mathbb{R}^{D\times 1}$ to represent the $D$-dimensional space of real-valued vectors. We also use $\|\cdot\|$ to denote the $L^2$-norm, which is an Euclidean norm. 

\section{Background} \label{sec:background}

\subsection{On-device AI (Distributed AI-to-AI) Communication}

On-device AI (distributed AI-to-AI) communication refers to a paradigm where AI models, \eg \ac{LM}, are embedded locally within devices and communicate with each other over a network. This decentralized approach allows devices to exchange complex, context-rich information, leveraging the capabilities of AI models to facilitate personalized and intelligent interactions without relying on centralized servers.

In the process of AI-to-AI communication using via \ac{LM}, the TX side initially encodes the input sequence.
This encoded data is then sent across through the physical layer to the RX side, where it is decoded back into its original form.
To enable this, several key components are required.
Both AI-Src-Enc at the TX and AI-Src-Dec at the RX must employ a shared tokenizer and embedder.
These tools play a pivotal role in accurately encoding and decoding the data.
Initially, the tokenizer divides the input sequence into discrete tokens. 
Subsequently, each token is transformed into an embedding representation through the embedder. 
This series of mapped embeddings is then input into the encoder to generate a transmittable representation. 
At the RX end, to reconstruct the original sequence from this representation, the same embedder is used to deduce the token index, which is then converted back to the corresponding token using the tokenizer.

The transmittable representation itself is a high-dimensional vector made up of floating point numbers. 
For its transmission via the physical layer, this vector needs to be converted into a form of bit-level information, suitable for digital transmission.
However, these floating-point numbers are particularly susceptible to bit-flips, which can occur due to noise within the physical layer as mentioned in \textbf{Q2}.
For instance, a 32-bit floating-point number representing 1.0 could change to infinity with just one bit-flip (bit error) in the second position, leading to a significant decline in transmission accuracy~\cite{lee2022seq2seq}.



\subsection{Vector Quantized Variational Autoencoders (VQ-VAE)}


\ac{VQ-VAE} are a version of variational autoencoders~\cite{kingma2013auto}, distinguished by their use of vector quantization techniques to effectively encode information into discrete latent representations~\cite{van2017neural}.
At the core, an autoencoder is a neural network that identifies latent spaces, which are complex and non-linear functions derived from the data. 
This neural network architecture is divided into encoder and decoder. 
The encoder's role is to process and transform the input data into a latent vector representation.
Following this, the decoder is tasked with the accurate reconstruction of the original data, using only the latent vector representation provided by the encoder.
Unlike traditional approaches where this latent representation is continuous~\cite{kingma2013auto}, the \ac{VQ-VAE} creates discrete latent representation by incorporating a discrete codebook.
This codebook, forming an array of vector entries each assigned a specific index, is utilized to discretize the latent space within the autoencoder.
By implementing this strategy, the \ac{VQ-VAE} facilitates the learning of a discrete latent space that effectively captures key features of the data.
The resultant latent representation in this model is a sequence of integer indices corresponding to the codebook entries, which not only allows for more efficient data compression but also maintains a level of reconstruction quality that is comparable to traditional methods.

\section{On-device AI Communication Systems} \label{sec:systemmodel}

\begin{figure*}[!ht]
\centering
\includegraphics[width=\linewidth]{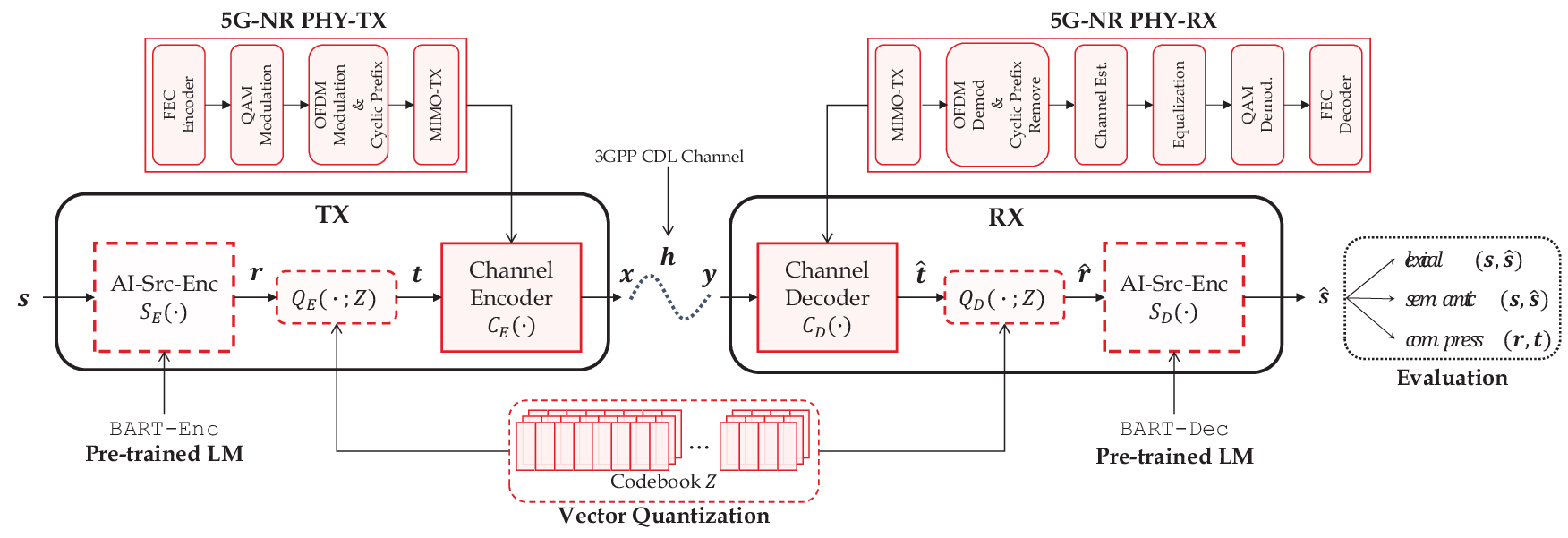} \\ 
\caption{\tblue{\textbf{Framework overview of on-device AI/LM \ac{PHY} communication systems} integrated with pre-trained language model.
This framework incorporates a link-level simulator to realistically emulate bit-level transmission within \ac{PHY} communication systems.
At the transmission (TX) end, the symbol stream $\vb* s$  undergoes AI-Src-Enc $S_{\mathrm{\mathbf{E}}}(\cdot)$, vector quantization $Q_{\mathrm{\mathbf{E}}}(\cdot; Z)$, and channel encoder $C_{\mathrm{\mathbf{E}}}(\cdot)$ to produce $\vb* x$.
Conversely, at the receiver (RX) end, the received signal $y$ is channel-decoded, vector-dequantized, and semantically decoded to recover the symbol $\vb*{\hat{s}}$.
We evaluate the system in three different criteria: lexical similarity, semantic similarity, and compression rate.}
}
\label{fig:framework}
\end{figure*}

\subsection{Problem Description}

\tblue{
Consider a sentence $\vb*{s}$ that maps to a symbol stream $\vb*{x}$ using the AI-Src-Enc $S_{\mathrm{\mathbf{E}}}(\cdot)$, vector quantization $Q_{\mathrm{\mathbf{E}}}(\cdot; Z)$, and channel encoder $C_{\mathrm{\mathbf{E}}}(\cdot)$ as follows:
\begin{equation}
{\vb*{x}} = C_{\mathrm{\mathbf{E}}} \left( Q_{\mathrm{\mathbf{E}}} \left( S_{\mathrm{\mathbf{E}}} \left( \vb*{s} \right); Z \right) \right),
\end{equation}
where $Z$ is the shared codebook. The symbol stream $\vb*{x}$ passes through a physical channel $h$ with noise $\vb*{n} \sim \mathcal{CN}(0, \sigma_n^2)$, resulting in the received signal $\vb*{y}$:
\begin{equation}
\vb*{y} = h \vb*{x} + \vb*{n}.
\end{equation}
At the receiver (RX) end, the received signal $\vb*{y}$ is processed by the channel decoder $C_{\mathrm{\mathbf{D}}}(\cdot)$, vector dequantizer $Q_{\mathrm{\mathbf{D}}}(\cdot; Z)$, and AI-Src-Dec $S_{\mathrm{\mathbf{D}}}(\cdot)$ to estimate the original sentence $\vb*{\hat{s}}$:
\begin{equation}
\vb*{\hat{s}} = S_{\mathrm{\mathbf{D}}} \left( Q_{\mathrm{\mathbf{D}}} \left( C_{\mathrm{\mathbf{D}}} \left( \vb*{y} \right); Z \right) \right).
\end{equation}

The architecture of the communication system is illustrated in Fig. \ref{fig:framework}, where the transmitter (TX) consists of the AI-Src-Enc $S_{\mathrm{\mathbf{E}}}(\cdot)$, vector quantizer $Q_{\mathrm{\mathbf{E}}}(\cdot; Z)$, and channel encoder $C_{\mathrm{\mathbf{E}}}(\cdot)$, and the receiver (RX) includes the channel decoder $C_{\mathrm{\mathbf{D}}}(\cdot)$, vector dequantizer $Q_{\mathrm{\mathbf{D}}}(\cdot; Z)$, and AI-Src-Dec $S_{\mathrm{\mathbf{D}}}(\cdot)$. Both TX and RX share the same codebook $Z$.

At the TX end, the sentence $\vb*{s}$ is encoded into a vector representation $\vb*{r}$ by $S_{\mathrm{\mathbf{E}}}(\cdot)$, mapped to discrete indices $\vb*{t}$ via $Q_{\mathrm{\mathbf{E}}}(\cdot; Z)$, and encoded into the symbol stream $\vb*{x}$ by $C_{\mathrm{\mathbf{E}}}(\cdot)$. This process adds redundancy to ensure reliable detection and correction of bit errors.

Conversely, at the RX end, the channel decoder $C_{\mathrm{\mathbf{D}}}(\cdot)$ decodes the received signal $\vb*{y}$, the vector dequantizer $Q_{\mathrm{\mathbf{D}}}(\cdot; Z)$ recovers the discrete indices $\vb*{\hat{t}}$ into vector representation $\vb*{\hat{r}}$, and the AI-Src-Dec $S_{\mathrm{\mathbf{D}}}(\cdot)$ decodes this to recover the estimated sentence $\vb*{\hat{s}}$.
}

\BfPara{On-device AI/LM = (New) Lossy Source Coding} \quad
While conventional (lossy) source coding focuses on compressing the source input by ensuring statistical similarity between the original and reconstructed signals ($\vb* s$ and $\hat {\vb* s}$), AI-Src-Enc takes a different approach.
The system aims to minimize lexical errors (\textit{i.e.,} lexical similarity) and semantic errors (\textit{i.e.,} semantic similarity) while also reducing the number of bits/symbols retrieved from AI-Src-Enc, thereby achieving compression.
In particular, it prioritizes compressing the embedding size (\ie the dimensionality of $\mathbf{r}$) while maintaining, or even enhancing, both lexical and semantic similarities between the original and reconstructed data.

While leveraging advanced traditional \ac{PHY} communication system's techniques (\eg \ac{MIMO}, \ac{OFDM}, channel coding, etc.) which prioritize achieving low \ac{BER} (or \ac{SER}), on-device AI/LM communication system focuses more on preserving the meaning between the original and reconstructed data to ensure successful distributed AI-to-AI (inter-AI) transmission.
To this end, we design and evaluate our \ac{E2E} on-device AI communication system within the context of bit-level \ac{PHY} communication transmission to see its \ac{E2E} performance in realistic scenarios.\footnote{
In this study, our primary focus is on \ac{PHY} layer functions, and we do not address MAC (or higher) layer functions, like \ac{ARQ} and \ac{HARQ}, which aim to ensure error-free transmission. It is important to note that in certain network scenarios—where simpler transmission is required, where transmitted data are latency-sensitive and cannot afford the delay caused by HARQ, or in broadcast situations where HARQ is not applicable, such as in Sidelink for device-to-device (D2D), URLLC, or LEO satellite networks—the retransmission process may be bypassed.
} Our rationale for this system is to preserve the meaning and context of the data, even in the presence of realistic channel noise and bit errors.

\subsection{System Model}



\BfPara{AI-Source-Encoder and AI-Source-Decoder} \quad
Both the AI-Src-Enc $S_{\mathrm{\mathbf{E}}}(\cdot)$ and AI-Src-Dec $S_{\mathrm{\mathbf{D}}}(\cdot)$ are constructed from a series of $6$ transformer layers~\cite{vaswani2017attention}. 
Each of these layers integrates a self-attention mechanism, positional encoding, and a densely connected layer, fortified by residual connections.
Notably, while they share a similar foundational architecture, the specific manner in which the self-attention operates diverges between the encoder and decoder components.
This distinction in the self-attention mechanism ensures specialized processing tailored to the unique demands of both encoding and decoding tasks.
The encoder employs a fully-visible self-attention strategy, granting the model the capability to focus on any token of the input while the decoder leverages an auto-regressive self-attention mechanism, limiting the model's attention solely to previous outputs, which is more appropriate for the streaming paradigm in which the decoder operates.
These architectures can be pre-trained on a large scale corpus by corrupting documents and computing the cross entropy loss between the decoder's output and the original document to learn the model generalizable knowledge~\cite{lewis-etal-2020-bart}.

Here, we employ such pre-trained checkpoints (\texttt{BART-base}~\cite{lewis-etal-2020-bart}) and use the encoder and decoder weights to initialize the weights of $S_{\mathrm{\mathbf{E}}}(\cdot)$ and $S_{\mathrm{\mathbf{D}}}(\cdot)$ respectively.
For a more detailed understanding of the en/decoder's operation, both $S_{\mathrm{\mathbf{E}}}(\cdot)$ and $S_{\mathrm{\mathbf{D}}}(\cdot)$ share the pre-trained embedding $\mathcal{E}$ and the pre-trained tokenizer $\mathcal{T}$.
Once the sentence $\vb* s$ is given to $S_{\mathrm{\mathbf{E}}}(\cdot)$, $\mathcal{T}$ tokenizes $\vb* s$ into tokens $\vb* s_{t} = [s_{t_1}, s_{t_2}, ... s_{t_n}]$ and maps each token to embedding $\vb* s_{e} = [s_{e_1}, s_{e_2}, ... s_{e_n}]$ by $\mathcal{E}$.
Then, $S_{\mathrm{\mathbf{E}}}(\cdot)$ encodes $\vb* s_{e}$ into hidden states $\vb* r = [r_1, r_2, ... r_n] \in \mathbb{R}^{n \times d_r}$ through multiple transformer layers, and passes it to channel encoder $C_{\mathrm{\mathbf{E}}}(\cdot)$.
In this context, $n$ denotes the total number of tokens, while $d_r$ specifies the dimension of the feature representation of each token.

After transmission is complete and the hidden states $\vb*{\hat{r}}$ are retrieved from the channel decoder $C_{\mathrm{\mathbf{D}}}(\cdot)$ and provided to $S_{\mathrm{\mathbf{D}}}(\cdot)$, the semantic decoder $S_{\mathrm{\mathbf{D}}}(\vb*{\hat{r}})$ establishes the conditional distribution $p_{\theta_{S_{\mathrm{\mathbf{D}}}}}\left(\vb*{\hat{s}}_i \mid \vb*{\hat{s}}_{0: i-1}, \vb*{\hat{r}}\right)$ and auto-regressively samples words from this distribution for each index.

\begin{figure}[t!]
\centering
\includegraphics[width=.95\linewidth]{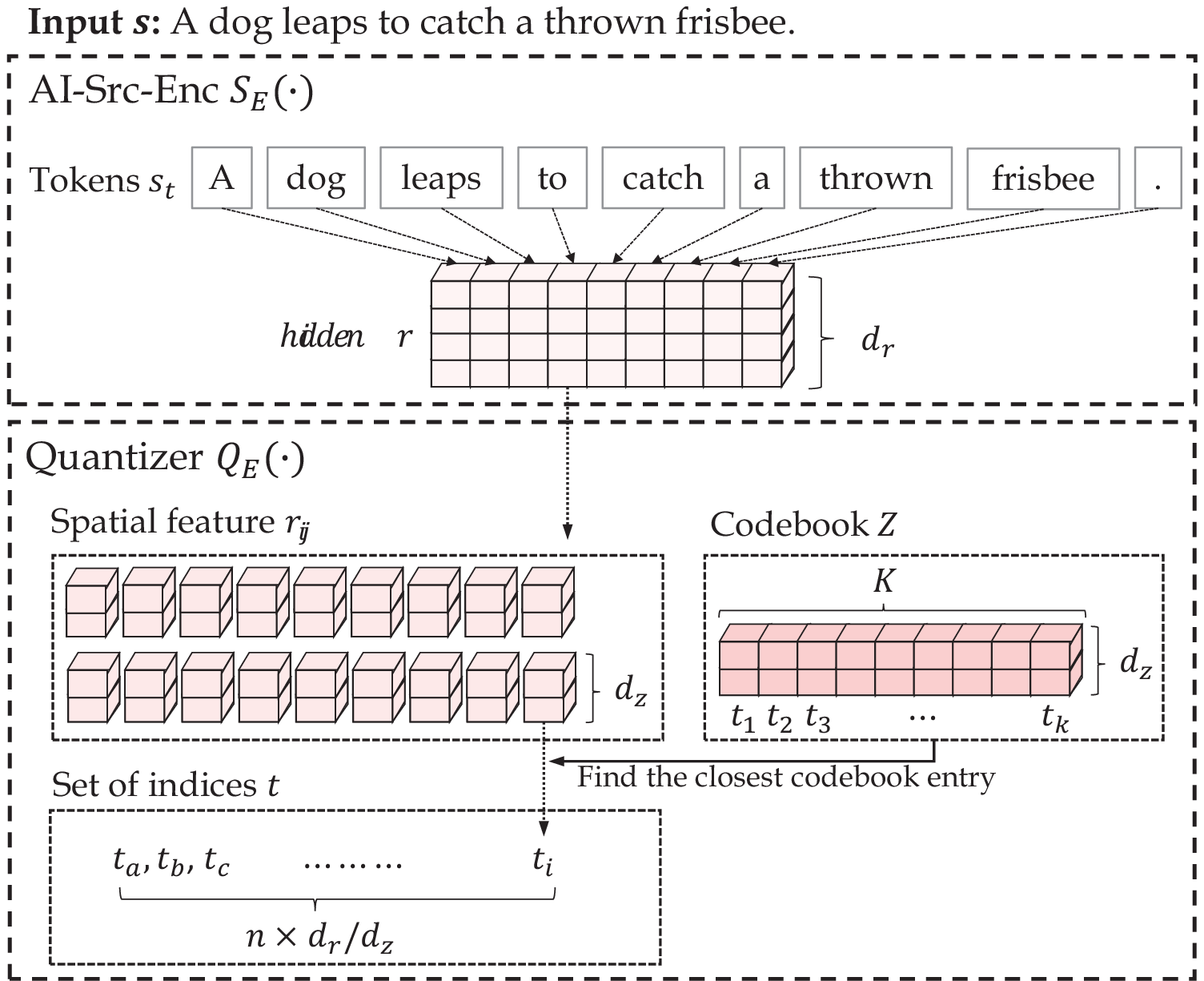} 
\caption{Architecture of AI-Source-Encoder (AI-Src-Enc) and Quantizer}
\label{fig:ai-src-enc}
\end{figure}

\BfPara{Vector Quantization (VQ-VAE)} \quad
\tblue{
Our framework employs vector quantization to transform continuous feature representations into discrete indices, enabling efficient data transmission. The core of this process is a discrete codebook 
\begin{equation}
Z=\left\{\boldsymbol{z}_k\right\}_{k=1}^K \in \mathbb{R}^{K \times d_z} ,     
\end{equation}
where $K$ denotes the codebook size and $d_z$ the dimension of each code vector. 

At the \ac{TX} end, the continuous feature vector $\vb* r \in \mathbb{R}^{n \times d_r}$, generated by the AI-Src-Enc $S_{\mathrm{\mathbf{E}}}(\cdot)$, is quantized into discrete indices $\vb* t \in \mathbb{R}^{n \times 1}$ using $Z$. This quantization aims for $d_z < d_r$ to ensure transmission efficiency. Specifically, each feature vector $\vb r_i \in \vb* r$ is divided into segments $\vb r_{i_j}$, each with dimension $d_z$. For instance, a feature vector $\vb r_i$ with dimension 768 and a codebook dimension of 256 can be segmented into three parts ($[\vb r_{i_1}, \vb r_{i_2}, \vb r_{i_3}]$), each of dimension 256. The quantizer $Q_{\mathrm{\mathbf{E}}}(\cdot; Z)$ maps each segment $\vb r_{i_j}$ to the closest codebook entry $\boldsymbol{z}_k \in Z$, forming the discrete representation $t_i \in \vb* t$:
\begin{equation}
t_i = Q_{\mathrm{\mathbf{E}}}(\vb r_i; Z) = \left(\underset{\boldsymbol{z}_k \in Z}{\operatorname{argmin}}\left\|\vb r_{i_j} - \boldsymbol{z}_k\right\|^2\right) \in \mathbb{R}^{(d_r/d_z) \times 1},
\end{equation}
where $(d_r/d_z)$ is the length of the encoded sequence. In the given example, $(d_r/d_z)$ equals 3.
}

Conversely, at the \ac{RX} end, the dequantizer $Q_{\mathrm{\mathbf{D}}}(\cdot; Z)$ converts the received discrete indices $\hat{t}_i$ back into continuous vectors $\hat{r}_i$ by directly referencing $Z$, reconstructing the feature representation $\hat{r}_i$ from $\hat{t}_i$.

\BfPara{\tpurp{Channel En/Decoder}} \quad
The channel encoder $C_{\mathrm{\mathbf{E}}}(\cdot)$ and decoder $C_{\mathrm{\mathbf{D}}}(\cdot)$ incorporate various \ac{PHY} communication functions, such as Polar code, QAM mapper, \ac{OFDM}, \ac{MIMO}, selected for their alignment with 5G-NR standards. This approach ensures that our framework mirrors real-world communication scenarios while acknowledging its partial compliance with 5G-NR protocols.

Firstly, the input bit for $C_{\mathrm{\mathbf{E}}}(\cdot)$ is grouped with other bits to form a codeword.
Polar codes, a form of linear block error correction codes, are considered one of the channel coding schemes in 5G-NR, where low complexity error correction is available~\cite{3GPP_NR_Polar}.
It adds redundancy to the codeword with a certain coderate $\rho$. This helps detect and correct bit errors introduced during transmission. The codeword is then further processed and segmented for transmission.

Each segmented codeword is mapped to a specific point on a constellation diagram, such as our considered QAM.
Higher QAM schemes offer more bits per symbol (better efficiency) but are more susceptible to noise.
The chosen QAM symbol is then represented by a combination of amplitude and phase variations in a carrier signal.
The number of bits per symbol used is denoted by $m$ and the mapper takes as input a message $u \in \{0,\dots,2^m-1\}$

The modulated symbols are divided into subcarriers across a wide spectrum.
This distributes the signal energy, making it more resilient to frequency-selective fading channels.
Each user (or subframe) can be modulated independently, enabling adaptive bit rate and channel equalization by using the received modulation and coding scheme (MCS) index.
The MCS index is dynamically selected and can be adjusted on a per-subframe basis, based on the latest channel quality information (CQI).

Data is transmitted simultaneously leveraging multiple antennas, \ie \ac{MIMO}, to exploit spatial diversity or multiplexing. This creates multiple independent channels, increasing reliability, data rate, or both. 
The processed signal is transmitted over the air.
At the receiver, the signal is received by multiple antennas. The received signal is then demodulated (QAM decoding) and demultiplexed (OFDM decoding). Channel coding algorithms (Polar decoding) are applied to correct any errors introduced during transmission. Finally, the decoded bits are reassembled into the original data.

\BfPara{Channel} \quad
In the on-device AI communication, the wireless propagation channel, henceforth simply referred to as \emph{channel}, introduces various types of noise and impairments, which can significantly affect the transmitted signal. These impacts include, but are not limited to, path loss, fading, Doppler shift, and \ac{AWGN}. 

3GPP has defined standardized models that are used for the simulation and testing of various wireless system concepts. These channel models describe time variation, delay dispersion, angular dispersion (at both link ends) and amplitude characteristics. 
We particularly leverage the standardized 3GPP \ac{CDL}-family channel models, such as CDL-A, CDL-B, and CDL-C~\cite{3gpp2018pathloss}. It is worth noting that while these CDL models are useful for testing, particularly for link-level simulation, they often diverge significantly from real-world channels.


\subsection{\tblue{Train Objectives}} \label{sec:loss}

\tblue{
Our framework is jointly trained using two loss functions: $\mathcal{L}_{\mathrm{CE}}$ and $\mathcal{L}_{\mathrm{VQVAE}}$.
The $\mathcal{L}_{\mathrm{CE}}$ minimizes the discrepancy between input sentence $\vb*{s}$ at the TX end and its predicted sentence $\vb*{\hat{s}}$ at the RX end, employing the cross entropy loss:
\begin{equation}
\begin{split}
    \mathcal{L}_{\mathrm{CE}}(\hat{\mathbf{s}}, \mathbf{s})=
    -\sum_{i=1}\nolimits
    p\left(\mathbf{s}_{t_i}\right) 
    \log 
    \left(p\left(\hat{\mathbf{s}}_{t_i}\right)\right)
    + \\
    \left(1-p\left(\mathbf{s}_{t_i}\right)\right) 
    \log 
    \left(1-p\left(\hat{\mathbf{s}}_{t_i}\right)\right) ,
\end{split}
\end{equation}
where $p\left(\mathbf{s}_{t_i}\right)$ is the true distribution in which the correct token at the $i$-th index has a probability of 1, and all other tokens have a probability of 0.
On the other hand, $p\left(\hat{\mathbf{s}}_{t_i}\right)$ denotes the predicted probability distribution over all the possible tokens for the $i$-th index.
$\mathcal{L}_{\mathrm{VQVAE}}$ aims to jointly train codebook $Z$ and the framework in end-to-end to bring the selected codebook vector ($\vb{t}$) as close as possible to the encoder output ($\vb{r}$) as follows:
\begin{equation}
\begin{array}{r}
\mathcal{L}_{\mathrm{VQVAE}}(\vb r, \vb t; Z)=\left\|\operatorname{sg}[\vb* r]-\vb* t\right\|_2^2 + \beta\left\|\operatorname{sg}[\vb* t]-\vb* r\right\|_2^2 .
\end{array}
\end{equation}
where $\operatorname{sg}[\cdot]$ represents the stop-gradient operation, ensuring no gradient is passed through and treating it as a constant that doesn't update.
It is applied to $\vb{r}$ to update only the codebook in the first term, while it is applied to $\vb{t}$ to update only the encoder output, thereby bringing these two closer together.
Our framework is trained by combining the two aforementioned loss functions as follows:
\begin{equation}
\begin{array}{r}
\mathcal{L}=\mathcal{L}_{\mathrm{CE}}(\hat{\mathbf{s}}, \mathbf{s}) + \mathcal{L}_{\mathrm{VQVAE}}( \vb r, \vb t; Z),
\end{array}
\label{eq:loss}
\end{equation}
In this training (fine-tuning) process, the trainable parameters include AI-Src-Enc $S_{\mathrm{\mathbf{E}}}(\cdot)$, AI-Src-Dec $S_{\mathrm{\mathbf{D}}}(\cdot)$, the embedding $\mathcal{E}$, quantizer $Q_{\mathrm{\mathbf{E}}}(\cdot; Z)$, dequantizer $Q_{\mathrm{\mathbf{D}}}(\cdot; Z)$, and the codebook $Z$.\footnote{
\tblue{
Wireless PHY components, including code rate \( \rho \), were not directly incorporated in the NN training. We acknowledge the potential for adaptive adjustment of specific standardized functions to improve robustness, which will be explored in future research.
}
}
}

        

\subsection{Optimization Techniques}
Our system employs three key optimization techniques to enhance performance:

\BfPara{Noise-Tuning} \quad
We optimize robustness against communication noise by adaptively tuning AI-Src-Enc and AI-Src-Dec to bit (or symbol) errors or channel noise. During fine-tuning, we expose them to simulated channel impairments (e.g., under the 3GPP CDL-A channel) alongside the channel encoder/decoder. This simple fine-tuning, without modifying any \ac{PHY} module, significantly improves reliability across diverse channel/communication conditions.


\BfPara{Codebook} \quad
To achieve efficient data compression, we utilize a \ac{VQ-VAE}. It learns a codebook of discrete vectors capturing the essence of the AI-Src-Enc's continuous representations. These codebook vectors, significantly smaller than the originals, are transmitted, reducing bandwidth demands.
Interestingly, it also mitigates the impact of bit errors during transmission as its representation learns the bit (or symbol) error pattern - potentially due to it learning the bit (or symbol) error pattern during pre-training. 
Upon reception, the \ac{VQ-VAE} decodes these indices back into high-dimensional vectors. This decoding utilizes the shared latent space (\ie $Z$) learned during pre-training, ensuring effective recovery of the semantic content within the compressed representation.
This method balances accuracy, compression, and inference complexity through the adjustable compression rate (and embedding size).

\BfPara{Pre-Training} \quad
To reduce the need for extensive application-specific training data, we leverage pre-trained models like \texttt{BART-base}. This pre-training equips the system with a broad base of generalizable knowledge, enhancing generalization performance.

\section{Experiments} \label{sec:exp}

\subsection{Dataset}
\BfPara{Train Dataset} \quad
Theoretically, for training purposes, any text can be utilized. We can set both the input $\vb*{s}$ and the output $\vb*{\hat{s}}$ to be identical for each sentence.
This approach aims to train the framework to ensure the faithful transmission of sentences without any modifications, aligning with the primary objective of the system.
However, following the precedent set by earlier studies in text semantic communication~\cite{Semantic_Geoffrey2021, hu2022one}, we utilize the European Parliament dataset~\cite{koehn-2005-europarl} (called \emph{EuroParl}) which is extracted from the proceedings of the European Parliament.

\BfPara{Test Dataset} \quad
For the evaluation, it is important to evaluate the generalization efficacy of framework on out-of-distribution data, particularly on sentences or even tokens not encountered during training.
To evaluate the generalizability, we randomly sample 1K sentences from the image-caption dataset \emph{Flickr}~\cite{young-etal-2014-image}.
The token distribution in this dataset differs from our training data, a result of reporting bias.
This bias emerges because people tend to report what interests them (\textit{e.g.,} parliamentary discussions) rather than typical and general facts (\textit{e.g.,} describing an image).

\subsection{Implementation Details}
\begin{table}[!h]   
\resizebox{.9\columnwidth}{!}{\begin{minipage}[t]{.9\columnwidth}
\centering
\caption{Type (or parameter) for channel en/decoder.}
\begin{tabular} {l l}
\toprule[1pt]
\textbf{Type (or Parameter)} & \textbf{Value} \\
\midrule[.8pt]
\emph{\textbf{Channel En/Decoder}} \\
\cmidrule(lr){1-1} \cmidrule(lr){2-2}
Channel coding & Polar code \\ 
Coderate ($\rho$) & $0.5$ \\
\# of coded bits & 960 \\
\# of information bits & 480 \\
Mapper & QAM \\ 
\# of bits per symbol ($m$) & $4$ \\
\# of OFDM symbols & 14 \\
\tblue{\# of \ac{TX}$\times$\ac{RX} antennas} & \tblue{2$\times$2 (1$\times$1)}\footnote{\tblue{Other than the results in Sec. V-B, which uses a 1x1 TX/RX configuration for a fair comparison, a MIMO (2x2) configuration is considered for the rest of the simulations.}} \\
Direction & Downlink \\
Carrier frequency & 2.6 [GHz] \\
Subcarrier spacing & 15 [kHz] \\
FFT size (\# of subcarrier) & 72 \\
Channel (train) & CDL-A, Rayleigh \\
Channel (test) & CDL-$\{$A$\sim$D$\}$, Rayleigh \\
\cmidrule(lr){1-2}
\emph{\textbf{AI-Source-En/Decoder}} \\
\cmidrule(lr){1-1} \cmidrule(lr){2-2}
Pre-trained LM ($\mathcal{T}$, $\mathcal{E}$, $S_{\mathrm{\mathbf{E}}}(\cdot)$, $S_{\mathrm{\mathbf{D}}}(\cdot)$) & \texttt{BART-base} \\
embedding dimension ($d_r$) & 768 \\
\cmidrule(lr){1-2}
\emph{\textbf{Codebook}} \\
\cmidrule(lr){1-1} \cmidrule(lr){2-2}
Dimension of each code ($d_z$) & 2 \\
Codebook size ($K$) & 1024 \\

\bottomrule[1pt]
\end{tabular}

\label{table_paramter_system}
\end{minipage}}
\end{table} 

Our entire framework is built using the Keras framework~\cite{chollet2015keras}.
For modeling various channel structures, we utilize the Sionna library~\cite{sionna}.
We employ the Huggingface library~\cite{wolf2020transformers} to load the pre-trained weights of the \texttt{BART-base} model, which serves as the foundation for initializing the embedding $\mathcal{E}$, tokenizer $\mathcal{T}$, AI-Src-Enc $S_{\mathrm{\mathbf{E}}}(\cdot)$ and the AI-Src-Dec $S_{\mathrm{\mathbf{D}}}(\cdot)$, respectively.
$\mathcal{T}$, pre-trained embedding $\mathcal{E}$.
\tblue{
In the fine-tuning phase, the Adam optimizer~\cite{KingBa15} is employed, with the sequence length capped at 64 tokens, and the training has been done for 3 epochs, using training loss in \eqref{eq:loss}\footnote{
\tblue{The pre-trained model may or may not use the same training loss function during its pre-training phase.}
}.
}
We conduct a grid search to find the best learning rate from the set [3e-4, 1e-4, 5e-5, 2e-5, 1e-5] and the batch size from [2, 4, 8], using the development dataset to determine the optimal parameters.
All of our experiments are run on an RTX 2080-Ti using 32-bit floating-point precision.

Detailed system parameters are summarized in Table \ref{table_paramter_system}.


\subsection{Evaluation} \label{sec:evaluation}
In traditional communication systems, the performance of information transmission is evaluated by measuring the accuracy of transmitting individual bits (0s and 1s), as reflected by the Bit Error Rate (BER), or by assessing the fidelity of conveying symbols, which are collections of bits, denoted by the Symbol Error Rate (SER).

In contrast, on-device AI communication prioritizes the transmission of meaningful content\footnote{This objective aligns with those of other research in the field of semantic communication.}, focusing on the efficient utilization of bandwidth for more effective data transfer. To accurately evaluate the performance of on-device AI communication systems, we utilize three specific metrics: lexical similarity (e.g., BLEU~\cite{papineni-etal-2002-bleu}), semantic similarity (e.g., SBERT~\cite{reimers-gurevych-2019-sentence}) between $\vb*{s}$ and $\vb*{\hat{s}}$, and the compression rate of $\rho$.

\begin{table*}[h!]
    \begin{minipage}{\linewidth}
	\centering
	\small      
        \resizebox{1.\linewidth}{!}
        {
            \centering
                \newcolumntype{R}{>{\raggedleft\arraybackslash}X}
\begin{tabular}{l|lccc|cccccc}
\toprule[1.5pt]
\multirow{2}{*}{\textit{\textbf{Framework}}} & \ \ \multirow{2}{*}{\textit{\textbf{\ac{PHY} Comm.}}} & \multirow{2}{*}{\textit{\textbf{Transmission}}} & \multirow{2}{*}{\textit{\textbf{Pre-Training}}} & \multirow{2}{*}{\textit{\textbf{Noise-Tuning}}} & \multicolumn{3}{c}{\textbf{EuroParl} (In-dist.)} & \multicolumn{3}{c}{\textbf{Flickr} (Out-dist.)} \\
\cmidrule(lr){6-8} \cmidrule(lr){9-11} 
& & & & & BLEU-3 & BLEU-4 & SBERT & BLEU-3 & BLEU-4 & SBERT \\
\midrule[.8pt] 
DeepSC~\cite{xie2021deep} & \ \xmark~(symbol-level) & \xmark & \xmark & \xmark & 0.6329 & 0.5802 & 0.7669 & 0.1878 & 0.1111 & 0.3737 \\
Seq2Seq-SC~\cite{lee2022seq2seq} & \ \cmark~(bit-level) & \texttt{Tanh} & \cmark & \xmark & 0.6428 & 0.5732 & 0.8142 & 0.5522 & 0.4795 & 0.7910 \\
\cmidrule(lr){1-11}
Ours ($1 \times 1$) & \ \cmark~(bit-level) & \texttt{VQ-VAE} & \cmark & \cmark & \bf 0.9959 & \bf 0.9946 & \bf 0.9959 & \bf 0.9714 & \bf 0.9682 & \bf 0.9998 \\
\bottomrule[1.5pt]
\end{tabular}
            \centering
        }
        \caption{Performance comparison study. Each baseline is trained with EuroParl training set and evaluated using both the EuroParl and Flickr test sets (Channel = Rayleigh, EbN0~$=7$ [dB]). The EuroParl test set assesses the in-distribution performance, whereas the Flickr test set evaluates the out-of-distribution performance.}
        \label{table:comparison}
    \end{minipage}
\end{table*}

\BfPara{BLEU} \quad
The BLEU score~\cite{papineni-etal-2002-bleu}, initially developed for evaluating machine translation, quantifies the correspondence between the $n$-grams of a generated sentence $\vb*{\hat{s}}$ and those in a reference sentence $\vb* s$~\cite{Semantic_Geoffrey2021, hu2022one}.
Here, $n$-grams mean a collection of $n$ successive words in a sentence.
BLEU score involves two key factors:
(1) $n$-gram-based precision of the generated sentence $\vb*{\hat{s}}$ and the reference sentence $\vb* s$ as follow:
\begin{equation}
p_n=\frac{\sum_{n \text {-gram} \in \vb*{\hat{s}}} \operatorname{Count}_{\text {clip}}(n \text {-gram})}{\sum_{n \text {-gram} \in \vb*{\hat{s}}} \operatorname{Count}(n \text {-gram})} .
\end{equation}
Here, $\operatorname{Count}_{\text {clip}}(n \text {-gram})$ represents the number of $n$-grams from the generated sentence $\vb*{\hat{s}}$ that are found in a reference sentence $\vb* s$ while $\operatorname{Count}(n \text {-gram})$ represents the total number of $n$-grams in the generated sentence $\vb*{\hat{s}}$.

However, simply counting identical $n$-grams can lead to an overestimation.
Consider the case where the reference sentence is ``I am a boy'' and the generated sentence is ``a a a a''.
In this scenario, the unigram precision $p_1$ would erroneously be computed as 1, as each occurrence of the unigram ``a'' is found in the reference sentence.

To address this issue, $\operatorname{Count}{\text {clip}}(n \text{-gram})$ is employed to cap the count at the highest frequency observed in the reference sentence $\vb* s$, thereby preventing overcounting.
In the given example, $\operatorname{Count}{\text {clip}}(\text{``a''})$ is adjusted to 1 instead of 4, reflecting the maximum occurrence of the unigram ``a'' in the reference sentence $\vb* s$;
(2) A brevity penalty is used to mitigate the influence of sentence length, preventing the overfitting to the sentence length.
This penalty comes into play when the length of  generated sentence $\vb*{\hat{s}}$ is shorter than the sentence length of the reference sentence $\vb* s$ as follow:
\begin{equation}
\mathrm{BP}= \begin{cases}1 , & \text { if }|\vb*{\hat{s}}|>|\vb* s| \\ e^{(1-|\vb* s| /|\vb*{\hat{s}}|)} . & \text { if }|\vb*{\hat{s}}| \leq|\vb* s|\end{cases}
\end{equation}

The BLEU score for $n$-grams, represented as BLEU-$n$, is calculated as the product of this brevity penalty and the exponential of the precision score for $n$-grams $\exp \left(\log p_n\right)$.
The overall BELU score is the brevity penalty multiplied by the exponential of the weighted sum of the log precision scores for different $n$-grams, represented as $\exp\left(\sum_{n=1}^N w_n \log p_n\right)$, where $w_n$ denotes the weight of the $n$-gram.
Here, BLEU-$n$ refers to the BLEU score considering only $n$-grams, and the general BLEU score is a weighted average of BLEU-1, BLEU-2, BLEU-3, and BLEU-4. 
Higher-order $n$-grams (\textit{i.e.,} longer $n$-grams) are indicative of the fluency and grammatical accuracy of the generated sentence $\vb*{\hat{s}}$.

\BfPara{SBERT} \quad
Despite a low lexical overlap between the sentences $\vb* s$ and $\vb*{\hat{s}}$, their semantic content may be closely aligned.
For example, the words \textit{``child"} and \textit{``children"} bear a close semantic relationship, yet a BLEU score based on lexical matching would not recognize this and would rate the similarity as zero.
To compute such semantic similarity, sentences can be represented into vector embeddings through an embedding model denoted as $\boldsymbol{M}$.
The degree of semantic similarity can then be quantified by computing the cosine similarity between these vector representations as follow:
\begin{equation}
\operatorname{match}(\hat{\mathbf{s}}, \mathbf{s})=\frac{\boldsymbol{M}(\mathbf{s}) \cdot \boldsymbol{M}(\hat{\mathbf{s}})^T}{\left\|\boldsymbol{M}(\mathbf{s})\right\|\left\|\boldsymbol{M}(\hat{\mathbf{s}})\right\|} .
\end{equation}

Existing research in semantic communication often employs BERT~\cite{devlin-etal-2019-bert} as the embedding model $\boldsymbol{M}$ to encode sentences.
To derive sentence embedding, they typically use methods like average pooling or extract the embedding from the first token (\textit{i.e.,} CLS pooling).
These embeddings are then used to calculate cosine similarity~\cite{Semantic_Geoffrey2021, hu2022one}.
Nonetheless, such pre-trained models, when not fine-tuned for the specific task of semantic textual similarity, may not effectively capture the true semantic nuances of sentences.
This is attributed to the anisotropic nature of the embedding space, which can lead to embeddings that do not properly represent the semantic variance across sentences~\cite{li-etal-2020-sentence}.
Here, we use SBERT~\cite{reimers-gurevych-2019-sentence}, which is fine-tuned on semantic textual similarity tasks, to encode the sentence embedding.



\BfPara{\tpurp{Compression Efficiency}} \quad
General (lossy) source coding in conventional systems emphasizes compact representation of information to enhance transmission/spectral efficiency, often at the expense of minor data loss. This contrasts with on-device AI communication, where the exchange of efficient representations prioritizes semantic fidelity over mere data volume reduction. The aim here is to transmit the essence and context of messages with minimal semantic distortion.

\tblue{
The compression rate in on-device AI communication, therefore, assesses how effectively AI-Src-Enc output is compressed. It is quantified by the ratio of the sizes of AI-Src-Enc output ($\mathbf{r}$) and channel encoder input ($\mathbf{t}$), described as:
\begin{equation}
\operatorname{compress}(\mathbf{r}, \mathbf{t})=\frac{\operatorname{size}(\mathbf{t})}{\operatorname{size}(\mathbf{r})},
\end{equation}
where $\operatorname{size}(\mathbf{x})$ denotes the size of vector $\mathbf{x}$.

To evaluate how compression impacts the conveyance of a message's meaning, we define compression efficiency ($\mathcal{E}_{\text{compress}}$) as a metric that examines the balance between the compression rate and semantic similarity:
\begin{equation}
\mathcal{E}_{\text{compress}} \triangleq \frac{\operatorname{match}(\hat{\mathbf{s}}, \mathbf{s}; \rho)}{\log_2{(\operatorname{compress}(\mathbf{r}, \mathbf{t}))}},
\end{equation}
where $\operatorname{match}(\cdot, \cdot; \rho)$ assesses the SBERT for a given $\rho$ compress rate.
A large value of $\mathcal{E}_{\text{compress}}$ indicates high compression efficiency, meaning that the model effectively balances compression and semantic fidelity. Conversely, a small value signifies lower compression efficiency.
}



\section{Experimental Results} \label{sec:num}

\begin{table*}[!h]
\centering
\resizebox{1.3\columnwidth}{!}{\begin{minipage}[h]{1.28\columnwidth}
\centering
\begin{tabularx}{1\linewidth}{l l}
    \toprule[1pt]
    \textbf{Transmit $\vb*{s}$} & A man pushes a bicycle along the beach as the sun sets behind him. \\
    \midrule[.8pt]
    \textbf{Baseline} & \textbf{Receive $\hat{\vb*{s}}$}\\
    \cmidrule(lr){1-1} \cmidrule(lr){2-2}
    DeepSC~\cite{xie2021deep}  & a man between a pity along the bureau as the danish sets behind him.\\
    Seq2Seq-SC~\cite{lee2022seq2seq} & A man pushes a bicycle along the beach as sun sets behind him.\\
    Ours ($1 \times 1$) & A man pushes a bicycle along the beach as the sun sets behind him.\\
    \bottomrule[1pt]
\end{tabularx}


\end{minipage}}
\caption{Received output examples of baselines and ours (Channel = Rayleigh, EbN0~$=7$ [dB]).}
\label{table:exampleSentences}
\end{table*}
\begin{figure*}[h!]
    \begin{subfigure}[t]{.24\linewidth}
        \centering  
        \includegraphics[width=\linewidth]{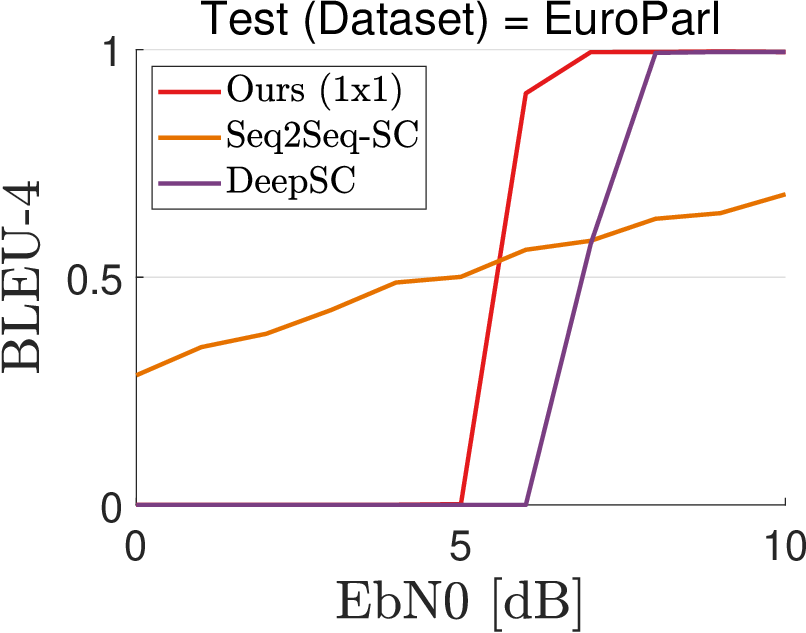}
        \caption{BLEU-4 (EuroParl)}
    \end{subfigure}\hspace*{.015\textwidth}%
    \begin{subfigure}[t]{.24\linewidth}
        \centering
        \includegraphics[width=\linewidth]{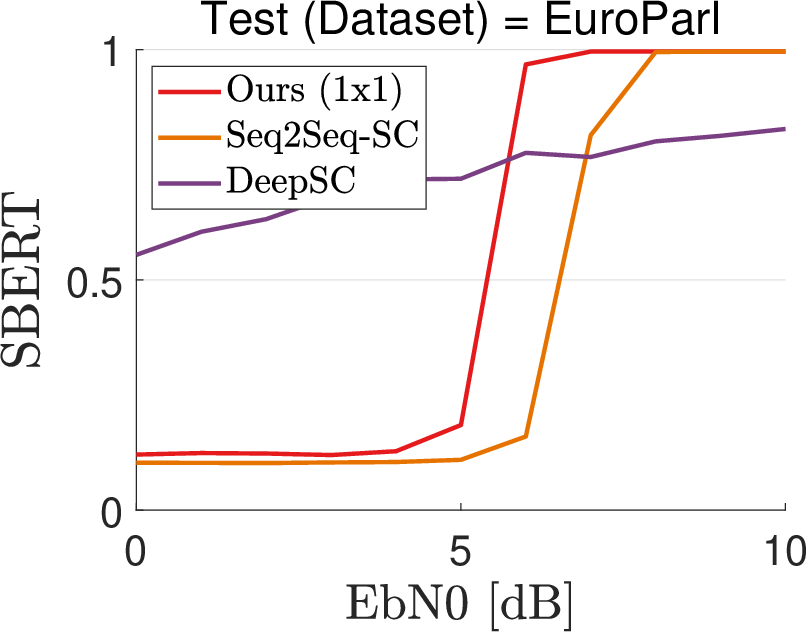}
        \caption{SBERT (EuroParl)}
    \end{subfigure}
    \begin{subfigure}[t]{.24\linewidth}
        \centering  
        \includegraphics[width=\linewidth]{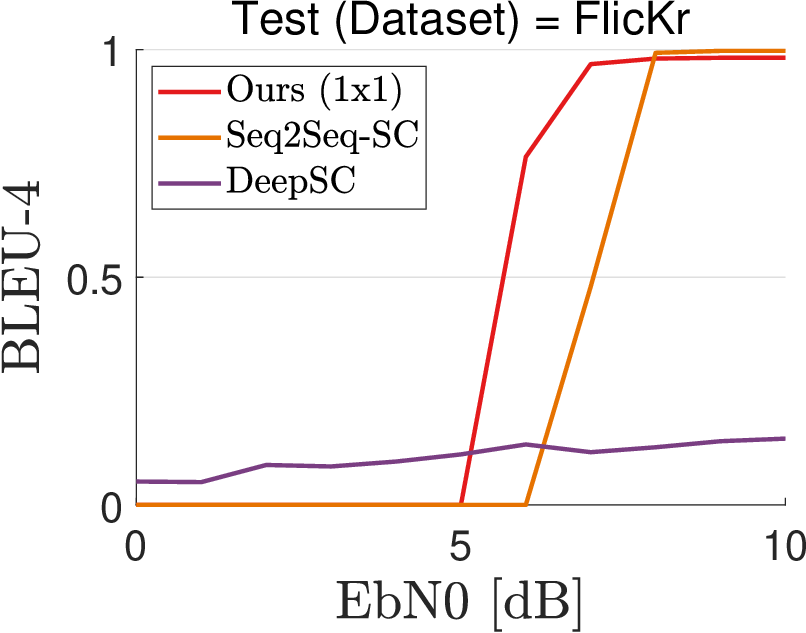}
        \caption{BLEU-4 (Flickr)}
    \end{subfigure}\hspace*{.015\textwidth}%
    \begin{subfigure}[t]{.24\linewidth}
        \centering
        \includegraphics[width=\linewidth]{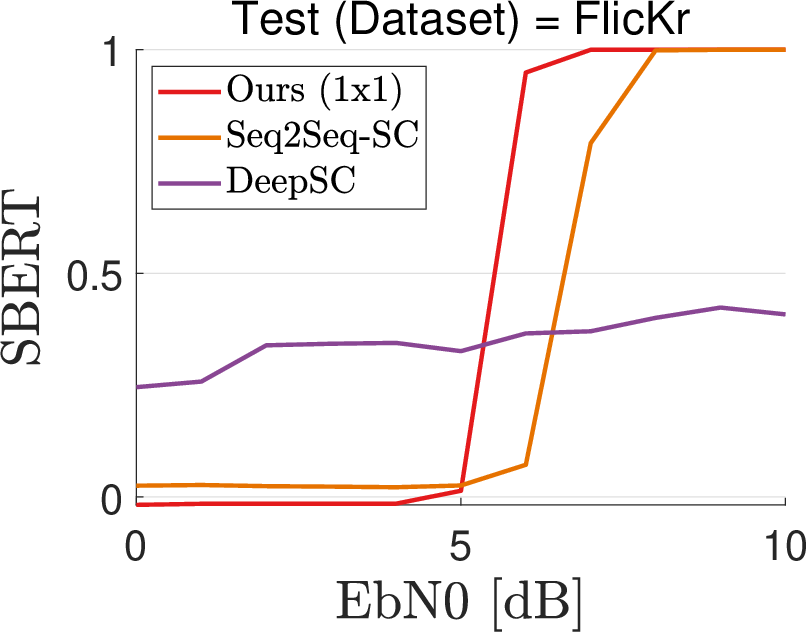}
        \caption{SBERT (Flickr)}
    \end{subfigure}
\caption{Performance comparison with existing frameworks under different EbN0 [dB] (Channel = Rayleigh)}
\label{fig:comparison}
\end{figure*}

\subsection{Setup/Baseline}
We primarily evaluate our proposed on-device AI/LLM communication frameworks (``Ours") from four distinct viewpoints:

\BfPara{\ac{PHY} Communication} \quad
\ac{PHY} communication conduct bit-level transmission by leveraging specific \ac{PHY} functions, \eg Polar coding, QAM mapper, \ac{OFDM}.
On the other hand, existing semantic communication works, exemplified by systems like DeepSC~\cite{Semantic_Geoffrey2021}), conduct symbol-level transmission, not considering the practical \ac{PHY} communication functions.

\BfPara{Pre-Training Utilization} \quad
(1) \texttt{w/o pre-training} utilizes a transformer architecture for the AI-Src-Enc and AI-Src-Dec, starting with random initialization; while
(2) \texttt{w/ pre-training} begins with pre-trained model weights to initialize the AI-Src-Enc and AI-Src-Dec.

\BfPara{Noise-Tuning Utilization} \quad
(1) \texttt{w/o noise-tuning} fine-tunes the framework in the absence of any simulated channel noise; while
(2) \texttt{w/ noise-tuning} incorporates simulated channel noise, inducing bit errors, into the fine-tuning process.
In particular, the systems is noise-tuned (fine-tuned) under the \ac{CDL} family of channel models and bit-level \ac{PHY} communication setup (see details in Table \ref{table_paramter_system}) and energy per-bit to noise ratio (EbN0)~$=5\sim15$ [dB]. 

\BfPara{Codebook Utilization} \quad
In \ac{PHY} communication, it is necessary to transform the vector output of the AI-Src-Enc into a discrete bit-level representation. 
One straightforward method is to directly map each floating-point number in the vector to its bit-level equivalent (\eg 768-dimensional vector results in 768$\times$32 bits).
However, this method is highly susceptible to channel noise, where even a single bit error can significantly distort the vector representation (\eg \emph{floating-point arithmetic error}).
To mitigate this issue, we explore and compare various techniques that are more robust to noise.
(1) \texttt{Direct} encodes the floating-point numbers of $\vb* r$ straight into bits for transmission;
(2) \texttt{Tanh} also transforms the floating-point numbers of $\vb* r$ into bits for sending but applies a hyperbolic tangent activation function to adjust for the cases when bit errors from channel noise push the received representation $\vb*{\hat{r}}$ beyond certain range (\eg $[ -1, 1 ]$).
By doing so, it realigns the out-of-bound values back into the $[ -1, 1 ]$ range, and has been used in Seq2Seq-SC~\cite{lee2022seq2seq};
(3) \texttt{VQ-VAE} employs vector quantization along with a codebook $Z$ to encode vector representation $\vb* r$ into a set of discrete codebook indices $\vb* t$ for transmission.

\subsection{Comparative Study}
We begin by evaluating the overall efficacy of our proposed framework against earlier works in semantic communication.
Each baseline is trained using the EuroParl training set and evaluated on both the EuroParl and Flickr test sets. There is a significant overlap in vocabulary between the EuroParl training and test sets, while the overlap is minimal between the EuroParl training set and the Flickr test set.  This setup allows for an assessment of the framework's generalizability.
For a fair comparison, all baseline models, including the baseline DeepSC which focuses on symbol-level transmission, are evaluated under identical conditions of $1 \times 1$ SISO and Rayleigh channel throughout the comparative study. 

Table~\ref{table:comparison} illustrates the performance comparison under a Rayleigh Channel with EbN0~$=7$ [dB]. Meanwhile, Table~\ref{table:exampleSentences} offers qualitative examples demonstrating the transmission accuracy of each methodology, and Fig.~\ref{fig:comparison} highlights the robustness of each approach.

\begin{table*}[h!t]
    \begin{minipage}{\linewidth}
	\centering
	\small      
        \resizebox{0.98\linewidth}{!}
        {
            \centering

\newcolumntype{R}{>{\raggedleft\arraybackslash}X}
\begin{tabular}{l|ccc|cccccc}
\toprule[1.5pt]
\multirow{2}{*}{\textit{\textbf{Combination}}} & \multirow{2}{*}{\textit{\textbf{Codebook}}} & \multirow{2}{*}{\textit{\textbf{Pre-Training}}} & \multirow{2}{*}{\textit{\textbf{Noise-Tuning}}} & \multicolumn{3}{c}{\textbf{EuroParl} (In-dist.)} & \multicolumn{3}{c}{\textbf{Flickr} (Out-dist.)} \\
\cmidrule(lr){5-7} \cmidrule(lr){8-10} & & & & BLEU-3 & BLEU-4 & SBERT & BLEU-3 & BLEU-4 & SBERT \\
\midrule[.8pt]
Ours & \texttt{VQ-VAE} & \cmark & \cmark & 0.9918 & 0.9888 & 0.9944 & 0.9296 & 0.9178 & 0.9959 \\
\cmidrule(lr){1-10}
\texttt{w/o pre-training} ($3$ epochs) & \texttt{VQ-VAE} & \xmark & \cmark & 0.0690 & 0.0393 & 0.3571 & 0.0002 & 0.0000 & 0.0080 \\
\texttt{w/o pre-training} ($10$ epochs) & \texttt{VQ-VAE} & \xmark & \cmark & 0.6136 & 0.5376 & 0.7277 & 0.0573 & 0.0246 & 0.0933 \\
\cmidrule(lr){1-10} 
\texttt{w/o noise-tuning} & \texttt{VQ-VAE} & \cmark & \xmark & 0.9515 & 0.9325 & 0.9804 & 0.9147 & 0.8907 & 0.9787 \\
\bottomrule[1.5pt]
\end{tabular}

            \centering
        }
        \caption{Performance variance w/ and w/o \texttt{pre-training} and \texttt{noise-tuning} (Channel = CDL-A, EbN0~$=4$ [dB]).}
        \label{table:ablation}
    \end{minipage}
\end{table*}

Tables~\ref{table:comparison} and \ref{table:exampleSentences} demonstrate that our framework significantly outperforms both DeepSC and Seq2Seq-SC. 
A key observation from Table~\ref{table:comparison} is the substantial decline in DeepSC's performance on the Flickr dataset, whereas the performance reduction for Seq2Seq-SC on Flickr is less severe in comparison. This difference can be mainly due to the use of \texttt{pre-training} which exhibits generalization capability for \ac{OOV} - that will be elaborated on a later section.

\tblue{
Fig.~\ref{fig:comparison} also demonstrates the robustness of our framework compared to Seq2Seq-SC across various EbN0 [dB].}
An interesting point in Fig.~\ref{fig:comparison} is the gradual increase in DeepSC's trend compared to the sharp rise in Seq2Seq-SC and our framework.
This primarily stems from whether the transmission takes place through bit-level transmission via practical \ac{PHY} communication.
Transmitting through the \ac{PHY} process information in bit-level, which requires converting the floating-point vector into bits, where bit errors caused by noise can lead to substantial changes in the information (called floating-point arithmetic error).
Therefore, in conditions of substantial noise, such as when EbN0~$<5$ [dB], transmitting vector representations through the \ac{PHY} becomes challenging.
Conversely, DeepSC transmits representations based on symbol-level, where noise is directly added into the continuous vector representation, while not considering the practical issues in bit-level operation. This results in such a gradual increase in its performance trend as it does not experience any floating-point error.


\begin{figure*}[h!]
\centering
\begin{subfigure}[t]{.22\linewidth}
\centering
\includegraphics[width=\linewidth]{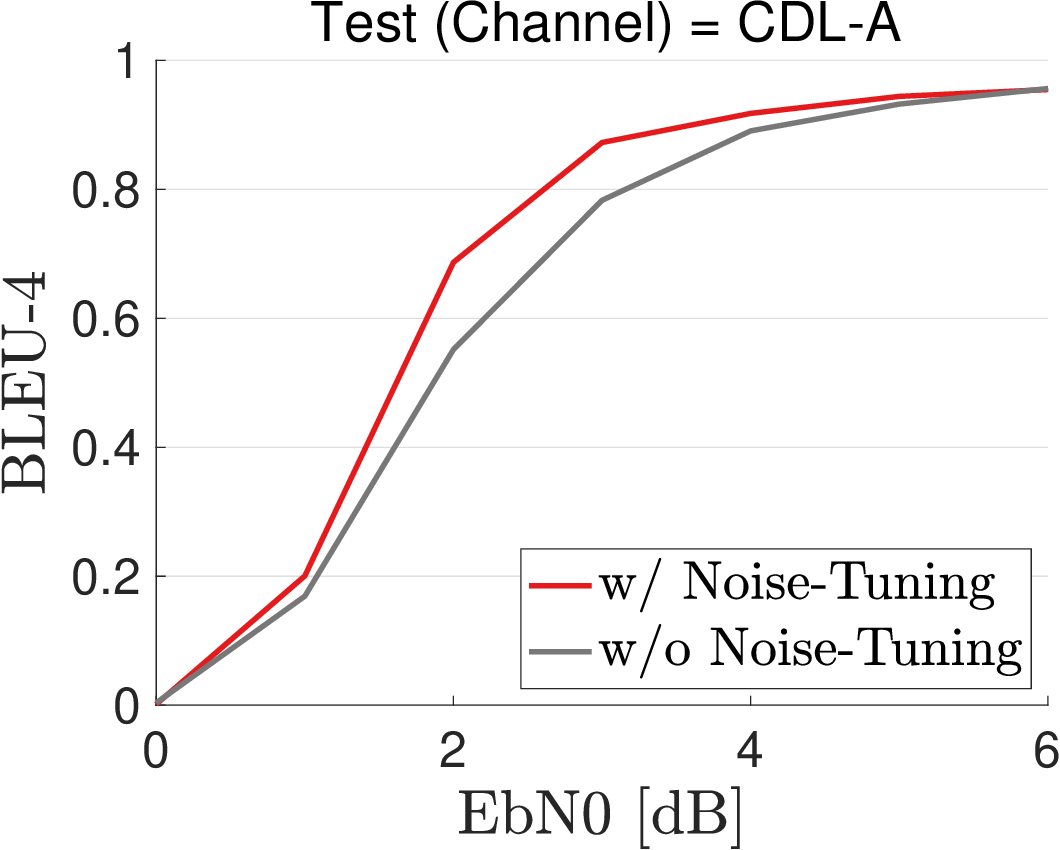}
\caption{BLEU-4 (In-dist.)}
\end{subfigure}\hspace*{.012\textwidth}%
\begin{subfigure}[t]{.22\linewidth}
\centering
\includegraphics[width=\linewidth]{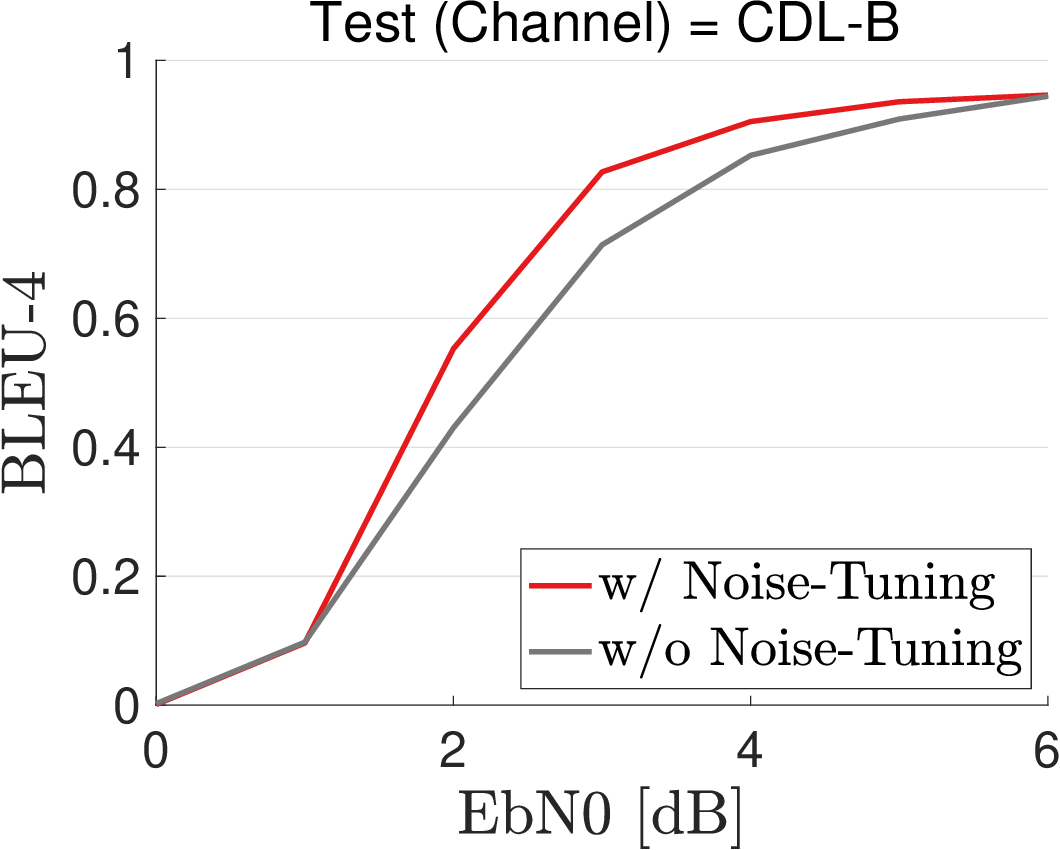}
\caption{BLEU-4 (Out-dist.)}
\end{subfigure}\hspace*{.012\textwidth}%
\begin{subfigure}[t]{.22\linewidth}
\centering
\includegraphics[width=\linewidth]{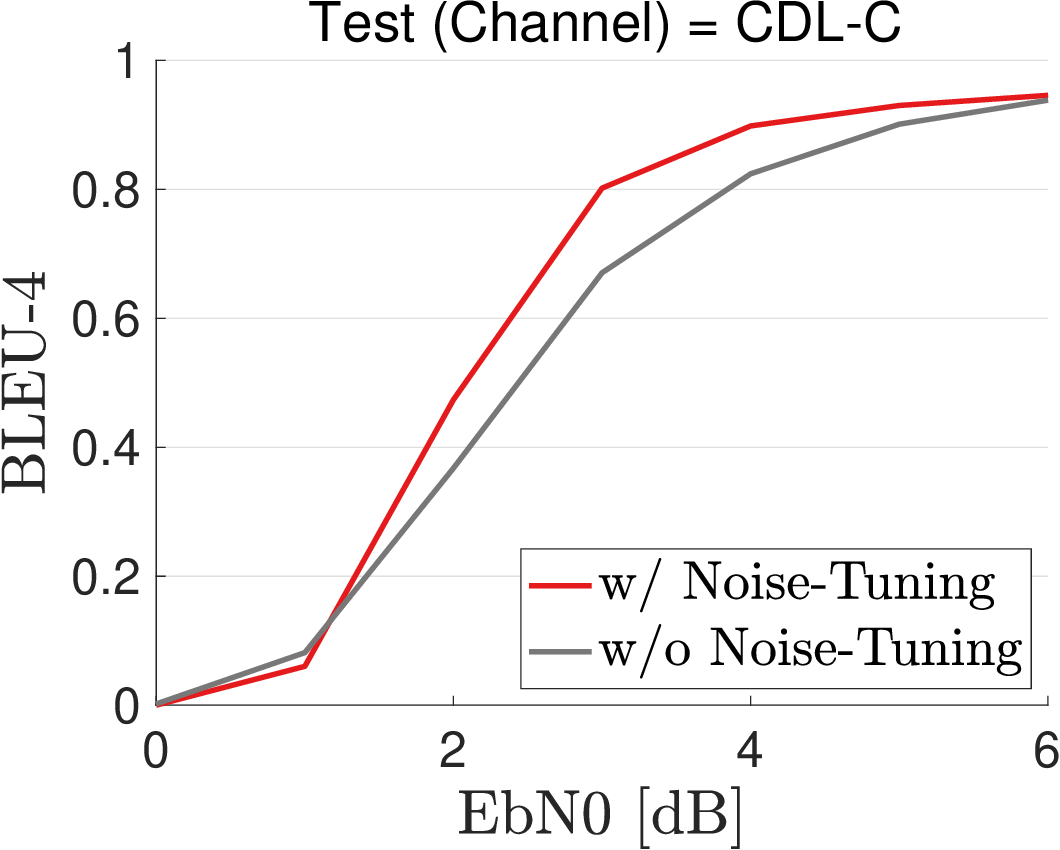}
\caption{BLEU-4 (Out-dist.)}
\end{subfigure}\hspace*{.012\textwidth}%
\begin{subfigure}[t]{.22\linewidth}
\centering
\includegraphics[width=\linewidth]{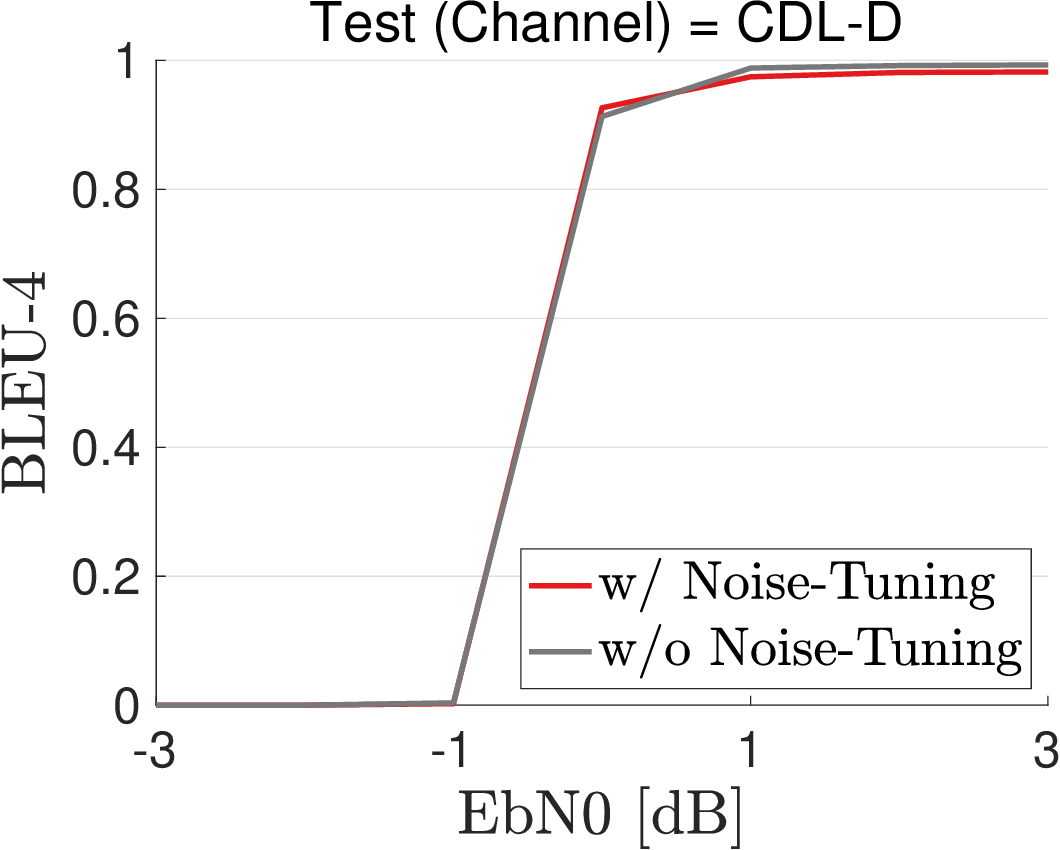}
\caption{BLEU-4 (Out-dist.)}
\label{fig:noisetune-d}
\end{subfigure}\vspace*{.02\textwidth} \\
\begin{subfigure}[t]{.22\linewidth}
\centering  
\includegraphics[width=\linewidth]{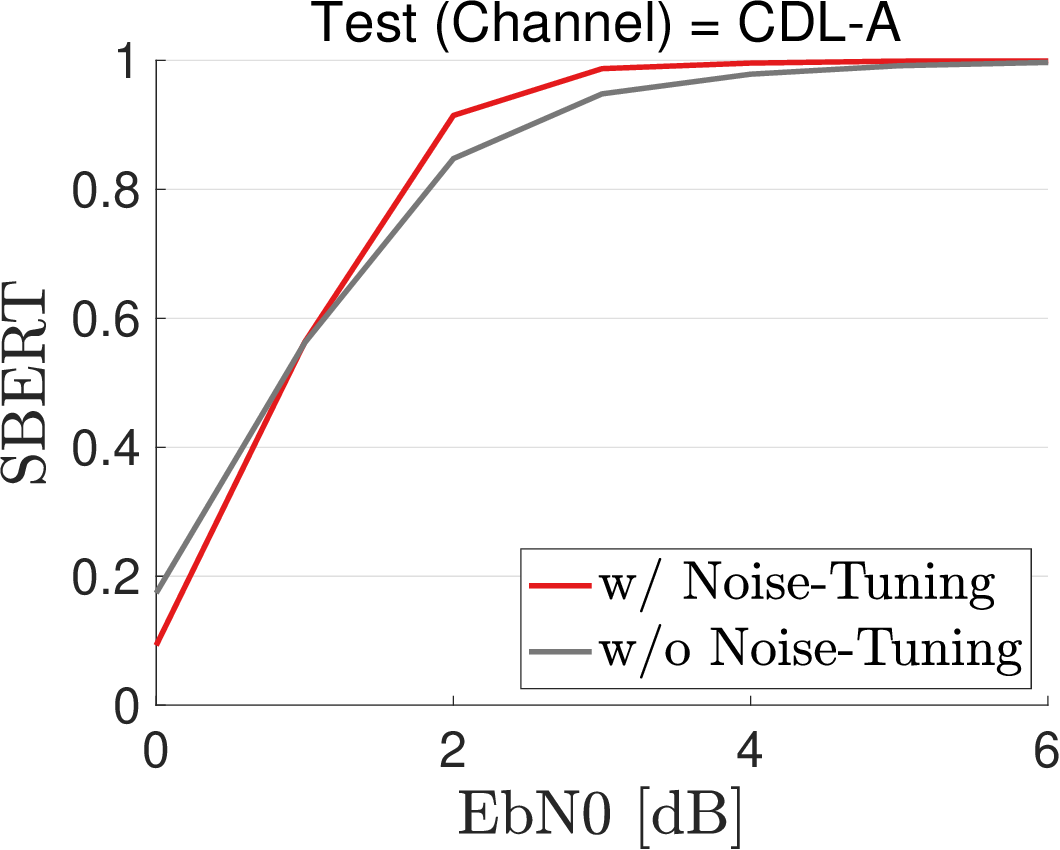}
\caption{SBERT (In-dist.)}
\end{subfigure}\hspace*{.012\textwidth}%
\begin{subfigure}[t]{.22\linewidth}
\centering  
\includegraphics[width=\linewidth]{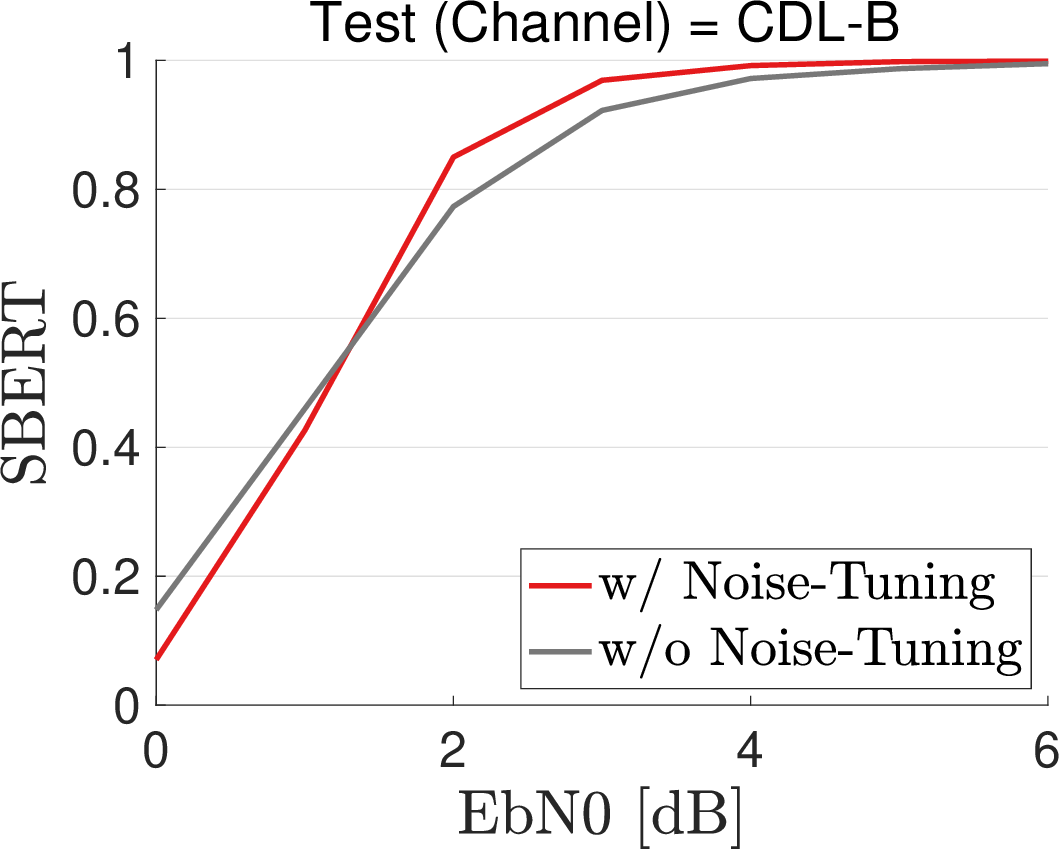}
\caption{SBERT (Out-dist.)}
\end{subfigure}\hspace*{.012\textwidth}%
\begin{subfigure}[t]{.22\linewidth}
\centering  
\includegraphics[width=\linewidth]{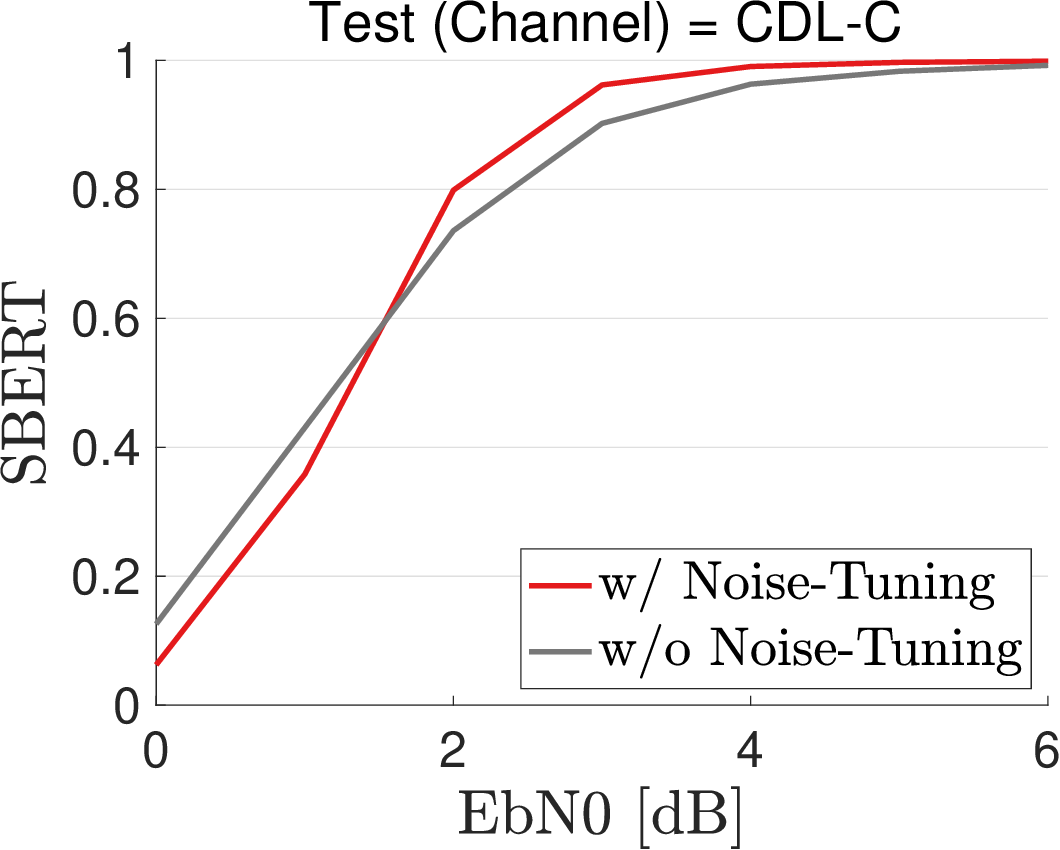}
\caption{SBERT (Out-dist.)}
\end{subfigure}\hspace*{.012\textwidth}%
\begin{subfigure}[t]{.22\linewidth}
\centering  
\includegraphics[width=\linewidth]{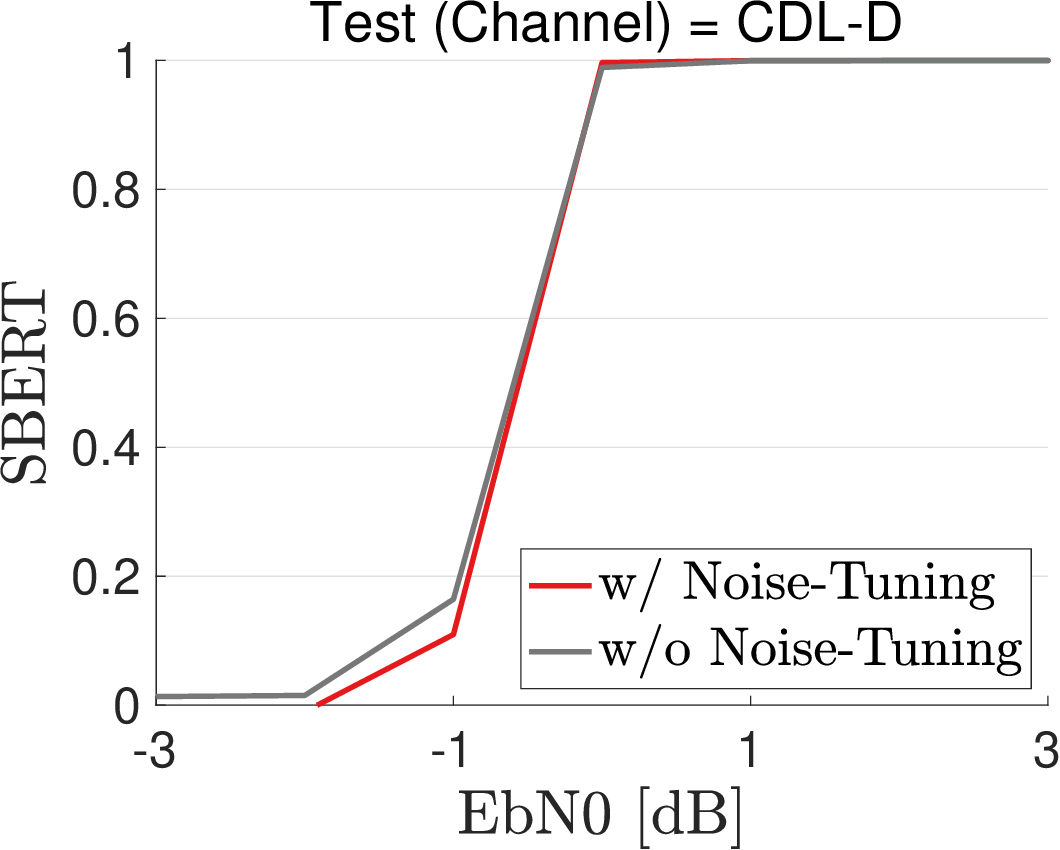}
\caption{SBERT (Out-dist.)}
\label{fig:noisetune-h}
\end{subfigure}
\caption{Impact of noise-tuning (w/o Noise-Tuning vs. w/ Noise-Tuning). The CDL-A channel assesses the in-distribution performance, whereas the CDL-B, -C, and -D evaluate the out-of-distribution performance.}
\label{fig:noisetune}
\end{figure*}

\subsection{Ablation Study}
In this part of the discussion, we explore the impact of each component within our optimized \ac{E2E} on-device AI communication framework and examine their respective roles.
Table~\ref{table:ablation} shows how performance varies with and without the use of \texttt{pre-training} and \texttt{noise-tuning}.
Additionally, Table~\ref{table:impact_codebook} shows the performance differences when using different transmission approaches including \texttt{Naive}, \texttt{Tanh} and \texttt{VQ-VAE}, showing the impact of the use of \texttt{Codebook}.

\subsubsection{Impact of Pre-Training}
As shown in Table~\ref{table:ablation}, \texttt{w/o pre-training} leads to a considerable performance drop.
After increasing the number of training epochs, the framework shows improvement on in-distribution data that shares a substantial vocabulary and distribution overlap with the training data.
However, its performance on other datasets with minimal overlap with the training data remains close to zero.
This indicates that \texttt{pre-training} plays a crucial role in the framework's ability to generalize - which allows this framework to be usable across a wide range of applications.

\subsubsection{Impact of Noise-Tuning}
As shown in Table~\ref{table:ablation}, \texttt{w/o noise-tuning} leads relatively small performance drop. However, it provides an intriguing insight that \ac{LM} and \ac{VQ-VAE} (\ie AI-Scr-Enc and AI-Src-Dec) are somewhat able to learn how to handle bit (or symbol) errors during their communication with another end. 

\begin{table}[!t]
\centering          
\resizebox{1.\linewidth}{!}{\begin{minipage}[h]{1.13\linewidth}
\centering
\newcolumntype{R}{>{\raggedleft\arraybackslash}X}
\begin{tabularx}{1\linewidth}{c | c | c c | c c }
\toprule[1.5pt]
\multirow{2}{*}{\textit{\textbf{EbNo ${[\mathrm{dB}]}$}}} & \textit{\textbf{Noise-Tuning}} & \textit{\textbf{Test}} & \multirow{2}{*}{\textit{\textbf{BER}}} & \multirow{2}{*}{\textit{\textbf{BLEU}}} & \multirow{2}{*}{\textit{\textbf{SBERT}}}  \\
& (\texttt{CDL-A}) &  (Channel) &  &  &  \\
\cmidrule[.8pt](lr){1-6}
\multirow{6}{*}{$2$~[dB]} & \xmark & \texttt{CDL-A} & \multirow{2}{*}{$2.124\times10^{-1}$} & 0.5801 & 0.8477  \\
 & \cmark & \texttt{CDL-A} &  & 0.7278 & 0.9147  \\
 \cmidrule(lr){2-6}
 & \xmark & \texttt{CDL-B} & \multirow{2}{*}{$2.332\times10^{-1}$} & 0.4588 & 0.7734  \\
 & \cmark & \texttt{CDL-B} & & 0.6125 & 0.8500  \\
 \cmidrule(lr){2-6}
 & \xmark & \texttt{CDL-C} & \multirow{2}{*}{$2.419\times10^{-1}$} & 0.4072 & 0.7359  \\
 & \cmark & \texttt{CDL-C} &  & 0.5394 & 0.7988  \\
\cmidrule[.8pt](lr){1-6} 
\multirow{6}{*}{$3$~[dB]} & \xmark & \texttt{CDL-A} & \multirow{2}{*}{$1.641\times10^{-1}$} & 0.7931 & 0.9481  \\
  & \cmark & \texttt{CDL-A} &  & 0.8893 & 0.9872  \\
  \cmidrule(lr){2-6}
 & \xmark & \texttt{CDL-B} & \multirow{2}{*}{$1.844\times10^{-1}$} & 0.7312 & 0.9224  \\
 & \cmark & \texttt{CDL-B} &  & 0.8441 & 0.9691  \\
 \cmidrule(lr){2-6}
 & \xmark & \texttt{CDL-C} & \multirow{2}{*}{$1.934\times10^{-1}$} & 0.6858 & 0.9023  \\
 & \cmark & \texttt{CDL-C} &  & 0.8260 & 0.9619  \\
\bottomrule[1.5pt]
\end{tabularx}
\centering
\end{minipage}}
\caption{Impact of noise-tuning (w/o Noise-Tuning vs. w/ Noise-Tuning).}
\label{table:noisetune}
\end{table}
\begin{table*}[h!]
    \begin{minipage}{\linewidth}
	\centering
	\small      
        \resizebox{0.9\linewidth}{!}
        {
            \centering
                \newcolumntype{R}{>{\raggedleft\arraybackslash}X}
\begin{tabular}{l|lcc|cccccc}
\toprule[1.5pt]
\multirow{2}{*}{\textit{\textbf{Case}}} & \ \ \ \multirow{2}{*}{\textit{\textbf{Codebook}}} & \multirow{2}{*}{\textit{\textbf{Pre-train}}} & \multirow{2}{*}{\textit{\textbf{Noise-tuning}}} & \multicolumn{3}{c}{\textbf{EuroParl} (In-dist.)} & \multicolumn{3}{c}{\textbf{Flickr} (Out-dist.)} \\
\cmidrule(lr){5-7} \cmidrule(lr){8-10} & & & & BLEU-3 & BLEU-4 & SBERT & BLEU-3 & BLEU-4 & SBERT \\

\midrule
\texttt{Naive} Transmission & \ \ \xmark~(\texttt{Naive}) & \cmark & \cmark & 0.0000 & 0.0000 & 0.0000 & 0.0000 & 0.0000 & 0.0000 \\
\texttt{Tanh} Transmission \tblue{(Seq2Seq-SC)} & \ \ \xmark~(\texttt{Tanh}) & \cmark & \cmark & 0.3660 & 0.2759 & 0.6691 & 0.1636 & 0.0983 & 0.5094 \\
\midrule
\texttt{VQ-VAE} Transmission \tblue{(Ours)} & \ \ \cmark~(\texttt{VQ-VAE}) & \cmark & \cmark & 0.9918 & 0.9888 & 0.9944 & 0.9296 & 0.9178 & 0.9959 \\
\bottomrule[1.5pt]
\end{tabular}

            \centering
        }
        \caption{\tblue{Performance variance with different transmission approaches (Channel = CDL-A, EbN0~$=4$ [dB]).}}
        \label{table:impact_codebook}
    \end{minipage}
\end{table*}

To assess the impact of noise-tuning in more detail, we present Table \ref{table:noisetune} and Fig. \ref{fig:noisetune}. These results compare the performance of AI-Src-Enc and AI-Src-Dec trained with and without noise-tuning (specifically, the CDL-A channel at EbN0 of $5\sim15$ [dB]) in four different communication conditions by testing those in CDL-A, B, C, and D channels. 
The performance difference showcases the impact of incorporating realistic communication errors into the fine-tuning process, indicating that AI-Src-Enc and AI-Src-Dec can learn and partially mitigate bit error patterns induced by noisy channel encountered during link-level \ac{PHY} communication.\footnote{
We conjecture that the AI-Src-Enc and AI-Src-Dec (\eg LM and VQ-VAE) learn bit error pattern of a pre-determined constellation in mapper/demapper. When symbol error occurs, neighbor symbols likely are detected, not far-distant symbols. During noise-tuning, such intelligent en/decoder adaptively modifies its $\mathcal{E}$ and $\mathcal{Z}$ to be more robust to bit errors.} However, how the AI-Src-Enc and AI-Src-Dec learn and overcome communication errors is still an open question that we defer to our future work.
Overall, the AI-Src-Enc/AI-Src-Dec archives EbN0 gain up to $\sim1$ [dB] by utilizing the noise-tuning.

An important challenge for on-device AI communication is their robustness to test-time distributional shifts that naturally occur when the test environment no longer matches the conditions in a training environment. This is present in wireless communication systems, where the propagation conditions may change whenever the user is moving. Thus, an important question is whether the on-device AI communication can retain its performance, both from the same communication condition in training (in-distribution) and other communication conditions (out-of-distribution). 
To this end, we also test its performance in CDL-B, -C, and -D channels, whose condition is not seen in training, in Table \ref{table:noisetune} and Figs. \ref{fig:noisetune}.\footnote{
CDL-A channels encompass both \ac{LoS} and \ac{NLoS} conditions, offering a more diverse representation of wireless channel propagation. In contrast, CDL-B and CDL-C channels are predominantly \ac{NLoS}. CDL-D channels, on the other hand, are characterized by \ac{LoS}. 
}
CDL-B and CDL-C results indicate that noise-tuning enhances the robustness even in channels not seen in training, while the CDL-D (Figs. \ref{fig:noisetune-d} and \ref{fig:noisetune-h}) case highlights that there is still room for improvement for the noise-tuning to be more channel-agnostic.


\begin{figure*}[h!]
\centering
\begin{subfigure}[t]{.19\linewidth}
\centering  
\includegraphics[width=\linewidth]{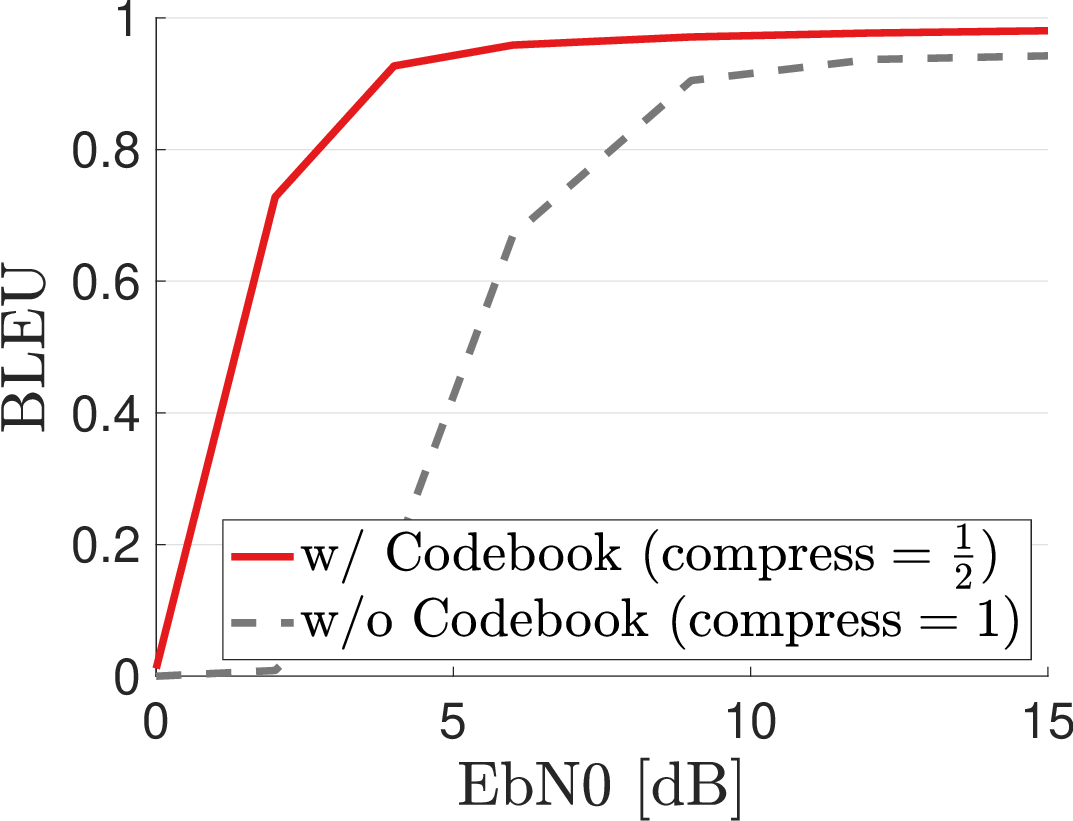}
\caption{w/ vs. w/o \texttt{VQ-VAE} (BLEU)}
\label{fig:vq-vae-a}
\end{subfigure}\hspace*{.01\textwidth}%
\begin{subfigure}[t]{.19\linewidth}
\centering  
\includegraphics[width=\linewidth]{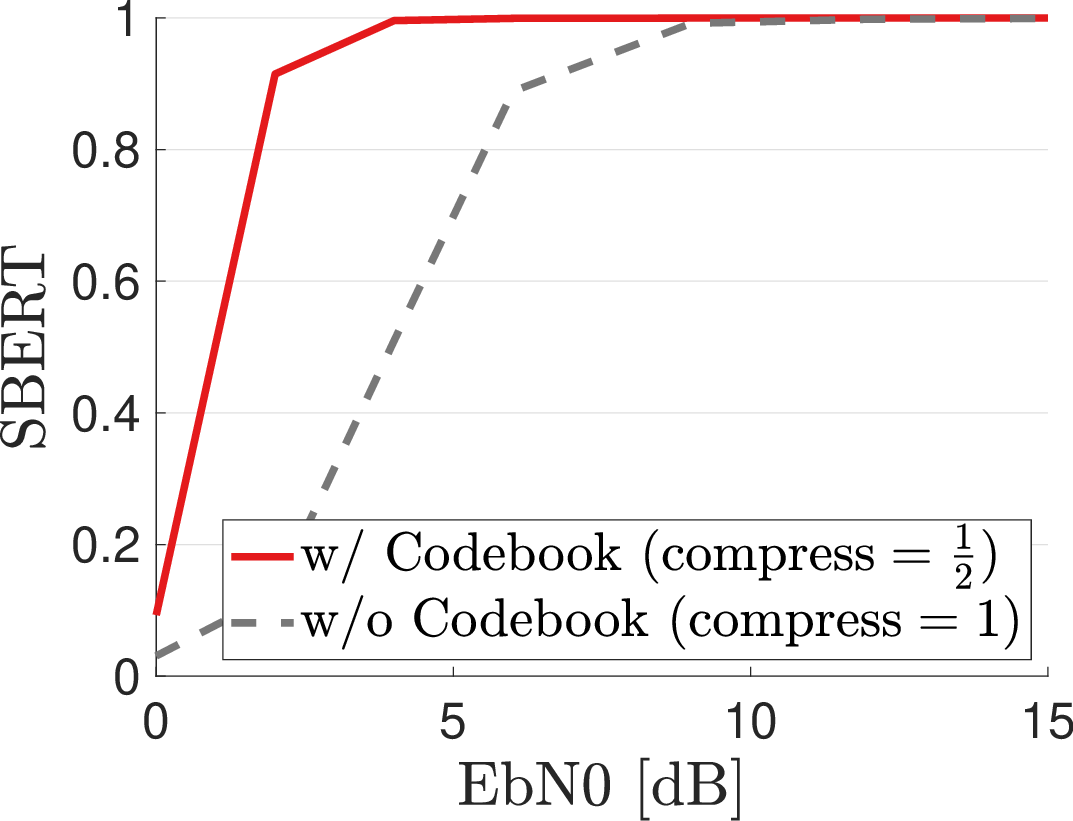}
\caption{w/ vs. w/o \texttt{VQ-VAE} (SBERT)}
\label{fig:vq-vae-b}
\end{subfigure}\hspace*{.01\textwidth}%
\begin{subfigure}[t]{.19\linewidth}
\centering
\includegraphics[width=\linewidth]{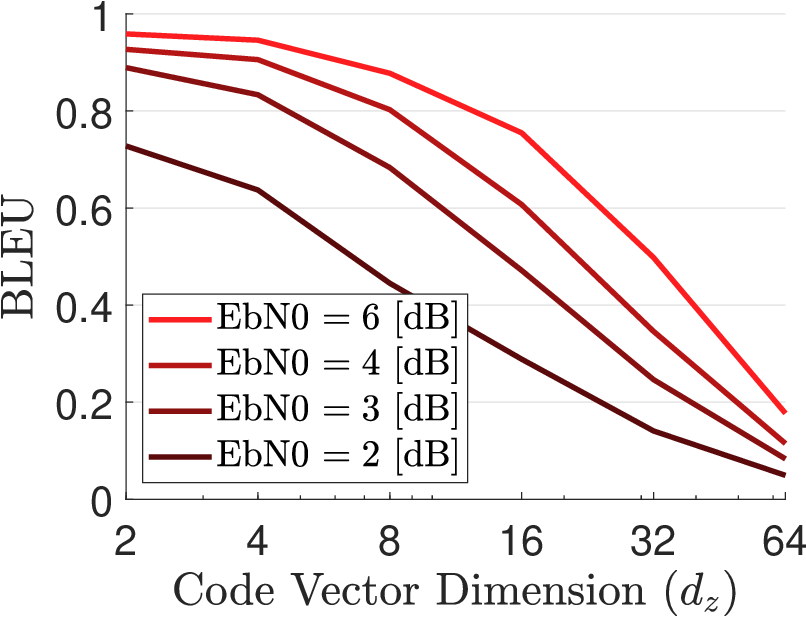}
\caption{Code vector dimension vs. Accuracy (BLEU)}
\label{fig:vq-vae-c}
\end{subfigure}\hspace*{.01\textwidth}%
\begin{subfigure}[t]{.19\linewidth}
\centering
\includegraphics[width=\linewidth]{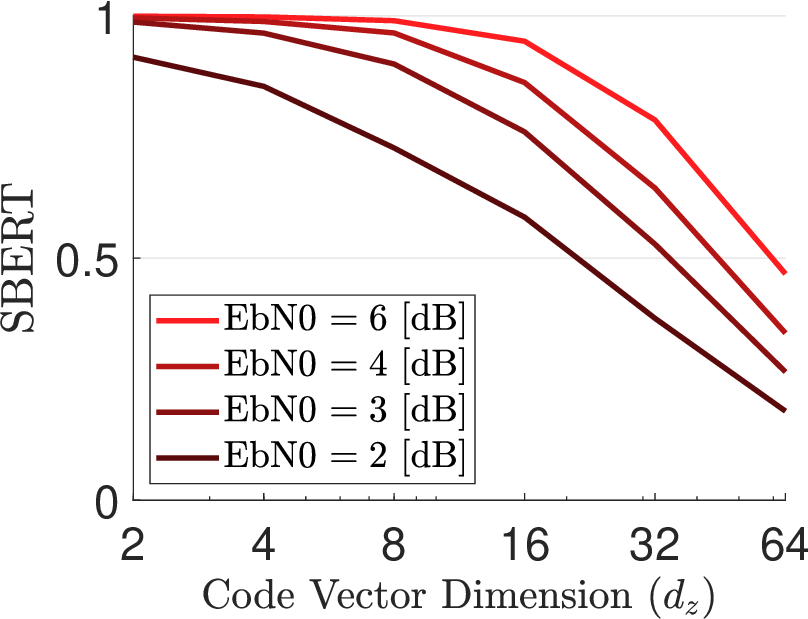}
\caption{Code vector dimension vs. Accuracy (SBERT)}
\label{fig:vq-vae-d}
\end{subfigure}\hspace*{.01\textwidth}%
\begin{subfigure}[t]{.19\linewidth}
\centering
\includegraphics[width=\linewidth]{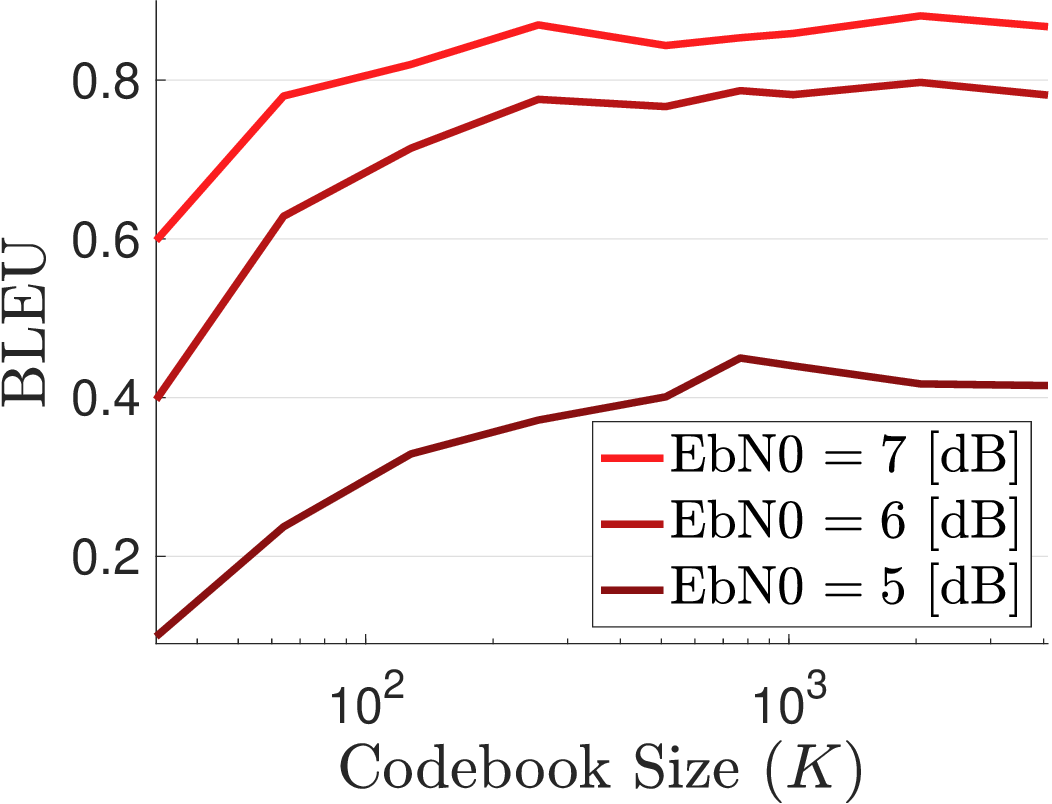}
\caption{Codebook size vs. Accuracy (BLEU)}
\label{fig:vq-vae_codebooksize}
\end{subfigure}
\caption{\tblue{Impact of codebook. Figs. \ref{fig:vq-vae-a} and \ref{fig:vq-vae-b} compares w/o Codebook and w/ Codebook. Figs. \ref{fig:vq-vae-c}-\ref{fig:vq-vae_codebooksize} show the tradeoff between compression and accuracy.}}
\label{fig:vq-vae}
\end{figure*}
\begin{figure}[h!]
\centering
\begin{subfigure}[t]{.45\linewidth}
    \centering  
    \includegraphics[width=\linewidth]{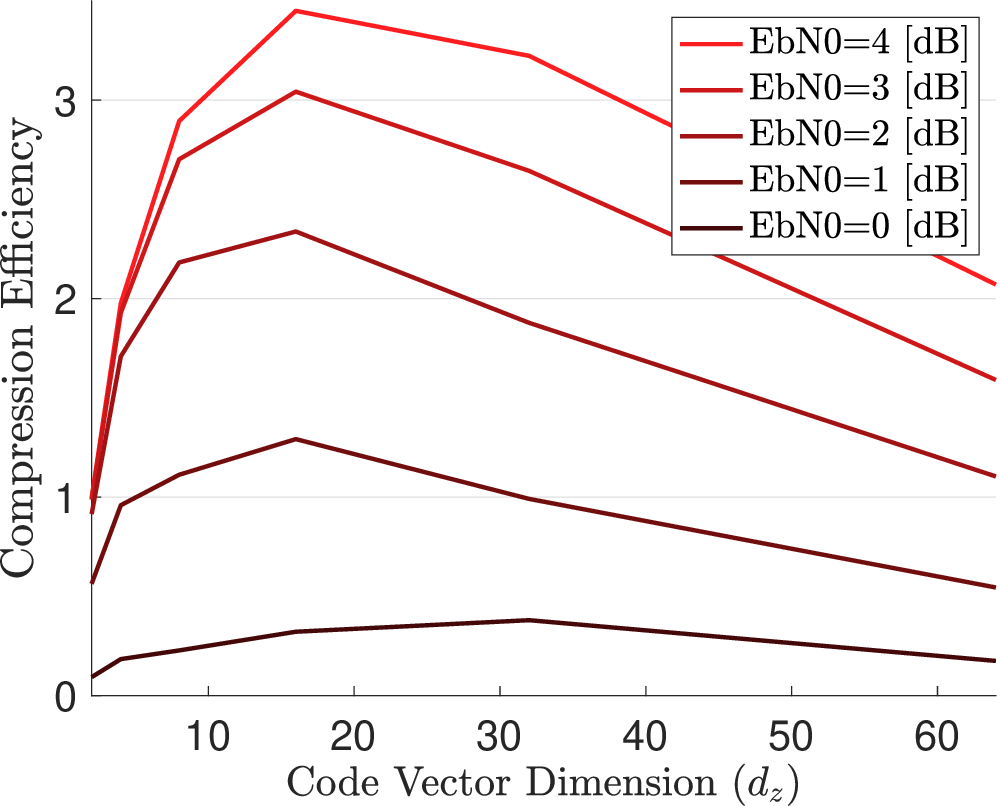}
\end{subfigure}
\caption{\tblue{Compression efficiency over compression rate}}
\label{fig:compression_efficiency}
\vspace{-1.em}
\end{figure}


\subsubsection{Impact of Codebook}
\tblue{
Table~\ref{table:impact_codebook} shows that the \texttt{VQ-VAE} transmission significantly outperforms other methods (\eg Seq2Seq-SC), achieving near-flawless results under the CDL-A channel at an EbN0~$=4$ [dB]. This demonstrates the effectiveness of using a discrete codebook for vector representation quantization in enhancing both accuracy and transmission robustness.

Figs. \ref{fig:vq-vae-a} and \ref{fig:vq-vae-b} further illustrate the impact of the codebook (\texttt{VQ-VAE}) over different EbN0 [dB].
In the scenario where a codebook is not used, the shape for transmitting information is an $n \times d_r$, where each element is a floating point vector.
Thus, the total size of the transmitted information $x$ becomes $n \times d_r \times 32$ bits, considering that each floating-point vector is 32 bits.
Conversely, when a codebook is used when $d_z=2$, the size of transmission is reduced to $n \times (d_r/2)$.
In this case, each point in the transmission is an integer index.
The total size of the transmitted information $x$ becomes $n \times (d_r/2) \times 32$, with the 32 bits now representing the size of each integer.
It is worth noting that using integer datatype is notably robust against rounding errors and precision loss; additionally, it allows for using smaller data types, such as 16-bit integers.
Remarkably, even with a $50 \%$ compression of the output of AI-Src-Enc, with-codebook consistently outperforms without-codebook, achieving EbN0 gains up to $\sim6$ [dB]. This suggests the potential of the encoded message from the AI-Src-Enc to be effectively compressed without losing its accuracy.


\BfPara{\tpurp{Tradeoff: Compression ($d_z$ and $K$) vs. Accuracy}}
The dimension of each code vector ($d_z$) and codebook size ($K$) in AI-Src-Enc/Dec impact the performance of the on-device AI communication systems.
Figs. \ref{fig:vq-vae-c} and \ref{fig:vq-vae-d} show how $d_z$ affects performance, examining increments through $\{2, 4, 8, 16, 32, 64\}$.
Increasing $d_z$ enhances compression but may reduce transmission accuracy, as compressing a data vector of size $d_z$ into a single index increases vulnerability to transmission errors—a single index error can impact the entire $d_z$.

Fig. \ref{fig:vq-vae_codebooksize} shows the relationship between $K$ and the performance.
Smaller $K$ reduces computational demands, yet it might cause an information loss during quantization, which can adversely affect the fidelity of the reconstructed representation.
On the other hand, increasing $K$ provides a more fine-grained resolution and greater accuracy.
However, this comes with increased computational cost due to the cost of training and inference overhead, and it may even hinder the convergence of training.

Additionally, Fig. \ref{fig:compression_efficiency} the relationship between compression efficiency ($\mathcal{E}_{\text{compress}}$) and code vector dimension, identifying a Pareto optimal point between the two.
Therefore, finding the right balance and designing an appropriate code vector dimension and codebook size is crucial for achieving a desirable balance between accuracy and efficiency. Thus, optimizing the compression rate (\eg $d_z$ and $K$) for AI-Src-Enc/Dec is crucial for striking an optimal balance between accuracy and computational efficiency.
}

\section{Conclusions} \label{sec:conclusion}
In this study, we integrated the \texttt{BART} pre-trained \ac{LM} with the \texttt{NVIDIA SIONNA} link-level simulator to simulate on-device AI communication within a 5G-NR framework. The performance of on-device communication was assessed in bit-level transmission across various \ac{PHY} communication setups, including 3GPP CDL-family channels, \ac{MIMO}, \ac{OFDM}, and Polar code.

Our analysis yielded several key insights from three optimization techniques: (1) Integration of pre-trained language models significantly enhances the generalization capabilities of the on-device AI communication system; (2) Noise-tuning results in an approximate EbN0 gain of around $1$ [dB], effectively improving performance in both familiar (CDL-A channel) and new (CDL-B$\sim$D channels) communication conditions; (3) Codebook utilization leads to an EbN0 gain of up to approximately $6$ [dB], accompanied by a $50 \%$ reduction in AI-Src-Enc output size. Even with a compression rate of $8$ (utilizing only $12.5 \%$ of the compressed information), the system maintains reasonable performance levels.

\tblue{
The implementation strategy involves two stages: 
\begin{itemize}
    \item \emph{Training Stage (Offline)}: The AI-Src-En/Dec embedding, along with the codebook, are trained by minimizing the loss function described in Sec. \ref{sec:loss}. This computationally intensive stage occurs offline, leveraging powerful resources. The training (fine-tuning) process can be optimized using efficient fine-tuning approaches, such as QLoRA \cite{dettmers2023qlora}.
    \item \emph{Inference Stage (On-Device)}: The on-device system employs the pre-trained AI-Src-Enc/Dec and codebooks, requiring only inference-level computations. The inference process can be optimized using on-device ML optimization techniques, such as pruning, $8$-bit quantization, or knowledge distillation.
\end{itemize}
}

\BfPara{Limitations} \quad
While our proposed framework showcases the performance of on-device AI communication under a bit-level transmission with a link-level \ac{PHY} communication setup, there are notable limitations that should be addressed to achieve full compliance with 5G-NR communication systems. Specifically, beyond the \ac{PHY}, the medium access control (MAC) or network layers typically employs automatic repeat request (ARQ) or hybrid ARQ (HARQ) protocols for packet error correction. The absence of consideration for these higher-layer protocols in our study limits the assessment of the practical performance of our framework. Incorporating ARQ or HARQ mechanisms into our simulation setup would provide a more comprehensive evaluation of the on-device AI communication framework's efficacy in real-world communication scenarios.

\BfPara{Open Questions} \quad
Experimental results in Sec. \ref{sec:evaluation} highlight that on-device AI can improve communication performance through its predictive and contextual learning abilities. While integrating channel encoding/decoding with AI-Src-Enc/Dec, such as through JSCC or AI-native \ac{PHY}/MAC, could unlock the full potential, this faces challenges like compatibility with 4G-LTE and 5G-NR, regulatory issues, and cost implications.

Therefore, employing AI as a complementary tool while keeping or carefully optimizing existing \ac{PHY} or higher-layer functions presents a more feasible approach. Future research could focus on: (1) finding the best noise-tuning conditions without altering \ac{PHY} functions; (2) exploring how language models can effectively counteract noisy conditions; (3) improving specific \ac{PHY} functions to boost AI communication; (4) determining if AI/LM can also learn packet error patterns to ease the HARQ (or ARQ) process.

\normalsize{
{\bf Acknowledgements:} This research was partly funded by the National Science Foundation (NSF). \\
We express our deep gratitude to Prof. Andreas F. Molisch for his invaluable feedback and multifaceted support throughout this work. We also thank Thomas Choi for his insightful discussions during the initial stages of our research.
}

\vfill\pagebreak

\bibliographystyle{IEEEtran} 
\bibliography{refs}

\begin{thebibliography}{10}
\providecommand{\url}[1]{#1}
\csname url@samestyle\endcsname
\providecommand{\newblock}{\relax}
\providecommand{\bibinfo}[2]{#2}
\providecommand{\BIBentrySTDinterwordspacing}{\spaceskip=0pt\relax}
\providecommand{\BIBentryALTinterwordstretchfactor}{4}
\providecommand{\BIBentryALTinterwordspacing}{\spaceskip=\fontdimen2\font plus
\BIBentryALTinterwordstretchfactor\fontdimen3\font minus \fontdimen4\font\relax}
\providecommand{\BIBforeignlanguage}[2]{{%
\expandafter\ifx\csname l@#1\endcsname\relax
\typeout{** WARNING: IEEEtran.bst: No hyphenation pattern has been}%
\typeout{** loaded for the language `#1'. Using the pattern for}%
\typeout{** the default language instead.}%
\else
\language=\csname l@#1\endcsname
\fi
#2}}
\providecommand{\BIBdecl}{\relax}
\BIBdecl

\bibitem{dhar2021survey}
S.~Dhar, J.~Guo, J.~J. Liu, S.~Tripathi, U.~Kurup, and M.~Shah, ``A survey of on-device machine learning: An algorithms and learning theory perspective,'' \emph{ACM Trans. Internet Things}, vol.~2, no.~3, Jul. 2021.

\bibitem{bommasani2022opportunities}
R.~Bommasani, D.~A. Hudson, E.~Adeli, R.~Altman, S.~Arora, S.~von Arx, M.~S. Bernstein, J.~Bohg, A.~Bosselut, E.~Brunskill \emph{et~al.}, ``On the opportunities and risks of foundation models,'' \emph{arXiv preprint arXiv:2108.07258}, 2022.

\bibitem{shijie2022recommendation}
S.~Geng, S.~Liu, and et~al., ``Recommendation as language processing ({RLP}): A unified pretrain, personalized prompt \& predict paradigm ({P5}),'' in \emph{Proc. ACM Conf. on Recommender Systems}, New York, NY, USA, 2022, p. 299–315.

\bibitem{cui2023survey}
C.~Cui, Y.~Ma, and et~al., ``A survey on multimodal large language models for autonomous driving,'' \emph{arXiv preprint arXiv:2311.12320}, 2023.

\bibitem{deletang2023language}
G.~Delétang, A.~Ruoss, P.-A. Duquenne, E.~Catt, T.~Genewein, C.~Mattern, J.~Grau-Moya, L.~K. Wenliang, M.~Aitchison, L.~Orseau, M.~Hutter, and J.~Veness, ``Language modeling is compression,'' \emph{arXiv preprint arXiv:2309.10668}, 2023.

\bibitem{lin2022overview}
X.~Lin, ``An overview of {5G} advanced evolution in {3GPP} release 18,'' \emph{IEEE Commun. Standards Mag.}, vol.~6, no.~3, pp. 77--83, 2022.

\bibitem{qin2021semantic}
Z.~Qin, X.~Tao, J.~Lu, W.~Tong, and G.~Y. Li, ``Semantic communications: Principles and challenges,'' \emph{arXiv preprint arXiv:2201.01389}, 2021.

\bibitem{bourtsoulatze2019deep}
E.~Bourtsoulatze, D.~B. Kurka, and D.~G{\"u}nd{\"u}z, ``Deep joint source-channel coding for wireless image transmission,'' \emph{IEEE Trans. on Cognitive Communications and Networking}, vol.~5, no.~3, pp. 567--579, 2019.

\bibitem{shao2021learning}
J.~Shao, Y.~Mao, and J.~Zhang, ``Learning task-oriented communication for edge inference: An information bottleneck approach,'' \emph{IEEE J. Sel. Areas Commun.}, vol.~40, no.~1, pp. 197--211, 2021.

\bibitem{huang2022toward}
D.~Huang, F.~Gao, X.~Tao, Q.~Du, and J.~Lu, ``Toward semantic communications: Deep learning-based image semantic coding,'' \emph{IEEE J. Sel. Areas Commun.}, vol.~41, no.~1, pp. 55--71, 2022.

\bibitem{tung2022deepwive}
T.-Y. Tung and D.~G{\"u}nd{\"u}z, ``{DeepWiVe}: Deep-learning-aided wireless video transmission,'' \emph{IEEE J. Sel. Areas Commun.}, vol.~40, no.~9, pp. 2570--2583, 2022.

\bibitem{wang2022wireless}
S.~Wang, J.~Dai, Z.~Liang, K.~Niu, Z.~Si, C.~Dong, X.~Qin, and P.~Zhang, ``Wireless deep video semantic transmission,'' \emph{IEEE J. Sel. Areas Commun.}, vol.~41, no.~1, pp. 214--229, 2022.

\bibitem{jiang2022wireless}
P.~Jiang, C.-K. Wen, S.~Jin, and G.~Y. Li, ``Wireless semantic communications for video conferencing,'' \emph{IEEE J. Sel. Areas Commun.}, vol.~41, no.~1, pp. 230--244, 2022.

\bibitem{xie2021deep}
H.~Xie, Z.~Qin, G.~Y. Li, and B.-H. Juang, ``Deep learning enabled semantic communication systems,'' \emph{IEEE Trans. on Signal Processing}, vol.~69, pp. 2663--2675, 2021.

\bibitem{han2022semantic}
T.~Han, Q.~Yang, Z.~Shi, S.~He, and Z.~Zhang, ``Semantic-preserved communication system for highly efficient speech transmission,'' \emph{IEEE J. Sel. Areas Commun.}, vol.~41, no.~1, pp. 245--259, 2022.

\bibitem{weng2023deep}
Z.~Weng, Z.~Qin, X.~Tao, C.~Pan, G.~Liu, and G.~Y. Li, ``Deep learning enabled semantic communications with speech recognition and synthesis,'' \emph{IEEE Trans. on Wireless Communications}, 2023.

\bibitem{vaswani2017attention}
A.~Vaswani, N.~Shazeer, N.~Parmar, J.~Uszkoreit, L.~Jones, A.~N. Gomez, {\L}.~Kaiser, and I.~Polosukhin, ``Attention is all you need,'' \emph{Advances in neural information processing systems (NIPS)}, vol.~30, 2017.

\bibitem{Semantic_Geoffrey2021}
H.~Xie, Z.~Qin, G.~Y. Li, and B.-H. Juang, ``Deep learning enabled semantic communication systems,'' \emph{IEEE Trans. on Signal Process.}, vol.~69, pp. 2663--2675, 2021.

\bibitem{hu2022one}
H.~Hu, X.~Zhu, F.~Zhou, W.~Wu, R.~Q. Hu, and H.~Zhu, ``One-to-many semantic communication systems: Design, implementation, performance evaluation,'' \emph{IEEE Commun. Lett.}, 2022.

\bibitem{brown2020language}
T.~Brown, B.~Mann, N.~Ryder, M.~Subbiah, J.~D. Kaplan, P.~Dhariwal, A.~Neelakantan, P.~Shyam, G.~Sastry, A.~Askell \emph{et~al.}, ``Language models are few-shot learners,'' \emph{Advances in neural information processing systems (NIPS)}, vol.~33, pp. 1877--1901, 2020.

\bibitem{srivastava2022beyond}
A.~Srivastava, A.~Rastogi, A.~Rao, A.~A.~M. Shoeb, A.~Abid, A.~Fisch, A.~R. Brown, A.~Santoro, A.~Gupta, A.~Garriga-Alonso \emph{et~al.}, ``Beyond the imitation game: Quantifying and extrapolating the capabilities of language models,'' \emph{arXiv preprint arXiv:2206.04615}, 2022.

\bibitem{thrun1998lifelong}
S.~Thrun, ``Lifelong learning algorithms,'' in \emph{Learning to learn}.\hskip 1em plus 0.5em minus 0.4em\relax Springer, 1998, pp. 181--209.

\bibitem{hoydis2022sionna}
J.~Hoydis, S.~Cammerer, F.~A. Aoudia, A.~Vem, N.~Binder, G.~Marcus, and A.~Keller, ``Sionna: An open-source library for next-generation physical layer research,'' \emph{arXiv preprint arXiv:2203.11854}, 2022.

\bibitem{van2017neural}
A.~Van Den~Oord, O.~Vinyals \emph{et~al.}, ``Neural discrete representation learning,'' \emph{Advances in neural information processing systems (NIPS)}, vol.~30, 2017.

\bibitem{lewis-etal-2020-bart}
M.~Lewis, Y.~Liu, N.~Goyal, M.~Ghazvininejad, A.~Mohamed, O.~Levy, V.~Stoyanov, and L.~Zettlemoyer, ``{BART}: Denoising sequence-to-sequence pre-training for natural language generation, translation, and comprehension,'' in \emph{Proc. the Association for Computational Linguistics}, Online, Jul. 2020, pp. 7871--7880.

\bibitem{lee2022seq2seq}
J.-H. Lee, D.-H. Lee, E.~Sheen, T.~Choi, and J.~Pujara, ``Seq2seq-sc: End-to-end semantic communication systems with pre-trained language model,'' \emph{Asilomar Conference on Signals, Systems, and Computers}, 2023.

\bibitem{kingma2013auto}
D.~P. Kingma and M.~Welling, ``Auto-encoding variational bayes,'' \emph{arXiv preprint arXiv:1312.6114}, 2013.

\bibitem{3GPP_NR_Polar}
{3GPP TS 38.212 v17.3.0}, ``{NR}; multiplexing and channel coding,'' Sep. 2022.

\bibitem{3gpp2018pathloss}
{3GPP}, ``Study on channel model for frequencies from 0.5 to 100 {GHz},'' \emph{Tech. Rep. 38.901 (V17.0.0)}, Mar. 2022.

\bibitem{koehn-2005-europarl}
P.~Koehn, ``{E}uroparl: A parallel corpus for statistical machine translation,'' in \emph{Proc. Machine Translation Summit X: Papers}, Phuket, Thailand, Sep. 2005, pp. 79--86.

\bibitem{young-etal-2014-image}
P.~Young, A.~Lai, M.~Hodosh, and J.~Hockenmaier, ``From image descriptions to visual denotations: New similarity metrics for semantic inference over event descriptions,'' \emph{Trans. of the Association for Computational Linguistics}, vol.~2, pp. 67--78, 2014.

\bibitem{chollet2015keras}
F.~Chollet \emph{et~al.}, ``Keras,'' \url{https://keras.io}, 2015.

\bibitem{sionna}
J.~Hoydis, S.~Cammerer, F.~{Ait Aoudia}, A.~Vem, N.~Binder, G.~Marcus, and A.~Keller, ``Sionna: An open-source library for next-generation physical layer research,'' \emph{arXiv:2203.11854 [cs.IT]}, Mar. 2022.

\bibitem{wolf2020transformers}
T.~Wolf, L.~Debut, V.~Sanh, J.~Chaumond, C.~Delangue, A.~Moi, P.~Cistac, T.~Rault, R.~Louf, M.~Funtowicz \emph{et~al.}, ``Transformers: State-of-the-art natural language processing,'' in \emph{Proc. Empirical Methods in Natural Language Processing}, 2020, pp. 38--45.

\bibitem{KingBa15}
D.~Kingma and J.~Ba, ``Adam: A method for stochastic optimization,'' in \emph{International Conference on Learning Representations (ICLR)}, San Diega, CA, USA, 2015.

\bibitem{papineni-etal-2002-bleu}
K.~Papineni, S.~Roukos, T.~Ward, and W.-J. Zhu, ``{BLEU}: a method for automatic evaluation of machine translation,'' in \emph{Proc. the Annual Meeting of the Association for Computational Linguistics}, Philadelphia, Pennsylvania, USA, 2002, pp. 311--318.

\bibitem{reimers-gurevych-2019-sentence}
N.~Reimers and I.~Gurevych, ``Sentence-{BERT}: Sentence embeddings using {S}iamese {BERT}-networks,'' in \emph{Proc. Empirical Methods in Natural Language Processing}, Hong Kong, China, Nov. 2019, pp. 3982--3992.

\bibitem{devlin-etal-2019-bert}
J.~Devlin, M.-W. Chang, K.~Lee, and K.~Toutanova, ``{BERT}: Pre-training of deep bidirectional transformers for language understanding,'' in \emph{Proc. Association for Computational Linguistics}, Minneapolis, Minnesota, 2019, pp. 4171--4186.

\bibitem{li-etal-2020-sentence}
B.~Li, H.~Zhou, J.~He, M.~Wang, Y.~Yang, and L.~Li, ``On the sentence embeddings from pre-trained language models,'' in \emph{Proc. the 2020 Conf. on Empirical Methods in Natural Language Processing}, Online, Nov. 2020, pp. 9119--9130.

\bibitem{dettmers2023qlora}
T.~Dettmers, A.~Pagnoni, A.~Holtzman, and L.~Zettlemoyer, ``Qlora: Efficient finetuning of quantized {LLM}s,'' \emph{arXiv preprint arXiv:2305.14314}, 2023.

\end{thebibliography}

\end{document}